\pdfoutput=1

\documentclass[aps,prd,twocolumn,superscriptaddress, nofootinbib]{revtex4}

\usepackage{mathrsfs}
\usepackage{amsmath}
\usepackage{amsfonts,amssymb}
\usepackage{graphicx}
\usepackage{graphics}
\usepackage[hidelinks]{hyperref}
\usepackage{bm} 
\usepackage[utf8]{inputenc}
\usepackage{abrev-jour-long}
\usepackage{tikz}

\usepackage{color}

\usepackage[normalem]{ulem}

\DeclareSymbolFont{symbolsC}{U}{pxsyc}{m}{n}
\DeclareMathSymbol{\coloneqq}{\mathrel}{symbolsC}{"42}

\begin{document}

\title[Caos em osciladores forçados]{Caos em osciladores forçados  sem amortecimento via mapas estroboscópicos
\\ \small
Chaos in undamped, forced oscillators via stroboscopic maps}

\author{Ronaldo S. S. Vieira}\email[]{ronaldo.vieira@ufabc.edu.br}

\author{Luiz H. R. Daniel}\email[]{luiz.daniel@aluno.ufabc.edu.br}

\affiliation{Centro de Ci\^encias Naturais e Humanas, Universidade Federal do ABC, 09210-580 Santo Andr\'e, SP, Brazil}

\author{Marcus A. M. de Aguiar}\email[]{aguiar@ifi.unicamp.br}

\affiliation{Instituto de F\'isica Gleb Wataghin, Universidade Estadual de Campinas, Unicamp 13083-970, Campinas, SP, Brazil}

\date{\today}

%
\begin{abstract}
A dinâmica não-linear não é um tema costumeiramente tratado em cursos de graduação em física. No entanto, sua importância dentro da mecânica clássica e da teoria geral de sistemas dinâmicos é inquestionável. Neste trabalho mostramos que esse assunto pode ser inserido na grade de um curso introdutório de mecânica clássica sem a necessidade de se desenvolver uma teoria robusta de dinâmica caótica. Para isso, tomamos como exemplos os osciladores não-lineares conservativos sujeitos a forças periódicas no tempo. Introduzindo o conceito de mapas estroboscópicos mostramos que é possível visualizar o aparecimento de caos nesses sistemas. Também abordamos o exemplo do pêndulo simples forçado aplicando o mesmo tratamento. Por fim, comentamos brevemente sobre a teoria mais geral de caos em sistemas hamiltonianos conservativos.
\newline \textbf{Palavras-chave:} dinâmica não-linear; caos; mapas estroboscópicos; seções de Poincaré.
\newline
\begin{center}
	\textbf{Abstract}
\end{center}

Non-linear dynamics is not a usually covered topic in undergraduate physics courses. However, its importance within classical mechanics and the general theory of dynamical systems is unquestionable. In this work we show that this subject can be included in the schedule of an introductory classical mechanics course without the need to develop a robust theory of chaotic dynamics. To do this, we take as examples conservative non-linear oscillators subject to time-dependent periodic forces. By introducing the concept of stroboscopic maps we show that it is possible to visualize the appearance of chaos in these systems. We also address the example of the forced simple pendulum applying the same treatment. Finally, we briefly comment on the more general theory of chaos in conservative Hamiltonian systems.
\end{abstract}


\maketitle
%
%
%
%
%

\section{Introdução}

Nas disciplinas de mecânica clássica da graduação em física é geralmente apresentada a resolução de equações diferenciais lineares (como por exemplo o oscilador harmônico amortecido forçado) \cite{watarimecanicaV1, symon1960mechanics}. O pro\-ble\-ma de pequenas oscilações, mesmo multidimensio\-nais, é também geralmente tratado apenas no caso linear \cite{marion2013classical, lemos2013mecanica}. Pro\-ble\-mas não-lineares, como em gravitação, são geralmente tratados apenas no caso esfericamente simétrico em que é possível resolver as equações exatamente \cite{marion2013classical, symon1960mechanics} ou por métodos perturbativos \cite{watarimecanicaV2}. Até mesmo a parte de mecânica analítica (lagrangiana e hamiltoniana) muitas vezes se resume à obtenção das equações de movimento específicas de um problema, mas sem entrar em detalhes das soluções dessas equações \cite{marion2013classical, symon1960mechanics}. 

Há um sentido nisso. Problemas não-lineares geralmente não possuem solução analítica em termos de funções elementares, com soluções obtidas apenas por quadraturas no caso unidimensional. Já no caso bidimensional, ou com forças externas dependentes do tempo, tais soluções são em geral impossíveis de se obter analiticamente (em\-bo\-ra as soluções existam, o que é garantido pelo teorema de existência e unicidade de equações diferenciais ordinárias \cite{boyce2010equaccoes, sotomayor1979li}). Entramos no domínio da \emph{dinâmica não linear}, que muitas vezes leva ao aparecimento de \emph{caos}. 

Tais assuntos são geralmente deixados para um curso de mecânica analítica em nível de pós-graduação \cite{lemos2013mecanica, deaguiarLivro, goldstein2002classical} ou para livros específicos de dinâmica caótica \cite{strogatz2018nonlinear, alligood2000bookchaos, ott2002chaos, tel2006chaoticBook, tabor1989chaos, lichtenbergLieberman1992, savi2006dinamica, fiedler1994caos}, que não estão contidos na grade curricular da maioria das universidades brasileiras. No entanto, a dinâmica caótica de sistemas tem sido estudada já há bastante tempo \cite{oestricher2007DiaClinNeuro, gleick2008chaos} e a pesquisa nesse campo de conhecimento tem crescido de maneira rápida, com aplicações em diversas áreas da física e astronomia, como por exemplo dinâmica galáctica \cite{binneytremaineGD} e astronomia dinâmica \cite{contopoulosOCDA2002}, relatividade geral \cite{hobill1994deterministic, saaVenegeroles1999PhLA, semerakSukova2010MNRAS, mosnaRodriguesVieira2022PRD}, dinâmica planetária \cite{theoryoforbits2, ferraz2021caos}, osciladores relativísticos \cite{kimLee1995PRE, vieiraMichtchenko2018CSF}, optomecânica \cite{aspelmeyerEtal2014RvMP}, mecânica quântica \cite{wimberger2014nonlinear, huang2018relativistic, novaes2021RBEF, tabor1989chaos, gubinSantos2012AmJP} e dinâmica de plasmas \cite{dasilvaEtal2002PhPl}, entre outras. Também tem bastante influência em ciências correlatas, como medicina \cite{oestricher2007DiaClinNeuro,skinner1992application} e biologia \cite{hastings1991chaos,maionchi2006chaos,vano2006chaos}. Os artigos e livros citados não caracterizam uma revisão dessa extensa área de pesquisa, mas sim exemplificam sua gama de aplicações. 

Vemos então a necessidade de um bacharel em física ter tido contato, mesmo que introdutório, com conceitos dessa área cujas aplicações são tão vastas. 
O público-alvo deste trabalho consiste do aluno que iniciou seus estudos de mecânica clássica, tendo tido contato com o oscilador harmônico forçado, com o pêndulo simples e com conceitos elementares qualitativos de dinâmica como o espaço de fases de uma partícula com um grau de liberdade e seu diagrama de fases; também pode ser útil como material didático complementar para que professores abordem o assunto em cursos de mecânica clássica. Desse modo, nosso objetivo aqui é mostrar que o tema pode ser apresentado de maneira intuitiva e quase ``na\-tu\-ral'' em um primeiro curso de mecânica clássica do bacharelado em física, de maneira a despertar o interesse por essa área de pesquisa tão abrangente nos dias atuais, e como consequência propor um caminho para a inserção desse tópico em cursos introdutórios de mecânica. Para isso trataremos de uma extensão que aparece ``naturalmente'' ao olharmos o oscilador harmônico como o movimento ao redor de um mínimo da energia potencial: osciladores anarmônicos forçados.

Vale ressaltar que outros artigos didáticos introdutórios em nível de graduação foram publicados sobre o tema, inclusive na Revista Brasileira de Ensino de Física \cite{weberEtal2023RBEF, deaguiar1994RBEF, moreira1993RBEF, cattaniEtal2016RBEF, oliveira2024RBEF}. Entendemos, no entanto, que a abordagem a\-pre\-sen\-ta\-da aqui é complementar aos artigos existentes e pode inclusive servir de motivação para o aprofundamento do leitor nessas outras referências, contribuindo para a literatura em português sobre o tema.

\section{Osciladores anarmônicos}
\label{sec:anarmonicos}

Vamos considerar o movimento unidimensional de uma partícula de massa $m$ em um campo potencial externo, sem amortecimento. Então a partícula está sujeita a uma energia potencial $U(x)$ vinda de sua interação com o campo externo. Dessa forma a equação de movimento para a partícula fica
\begin{equation}\label{eq:eqmovimentoU}
m\,\ddot x + \frac{dU}{dx} = 0
\end{equation}
e conservação da energia mecânica $E$ nos dá 
\begin{equation}\label{eq:Emec}
E = \frac{1}{2}\,m\dot{x}^2+ U(x)\, .
\end{equation}
Vale notar que o movimento unidimensional sob a ação de uma energia potencial é bem conhecido e apresentado nos livros básicos de mecânica clássica \cite{watarimecanicaV1, marion2013classical, symon1960mechanics}. Em par\-ti\-cu\-lar, para o movimento limitado, o diagrama de fases (o conjunto de órbitas da partícula no espaço $x-\dot{x}$) é formado geralmente por curvas fechadas \cite{marion2013classical}, cujo período de oscilação possui a expressão \cite{watarimecanicaV1}
\begin{equation}\label{eq:periodo}
T = 2\int_{x_{\rm{min}}}^{x_{\rm{max}}}\frac{dx'}{\sqrt{\frac{2}{m}\left[E - U(x')\,\,\right]}\,\,} \, ,
\end{equation}
onde $x_{\rm{min}}$ e $x_{\rm{max}}$ são os pontos de retorno do movimento, que dependem da energia mecânica $E$ do sistema.

Vamos também supor que $x=0$ é um ponto de equilíbrio (estável ou instável) do sistema, isto é, um mínimo ou máximo local da energia potencial, respectivamente. No caso em que $x=0$ é estável (mínimo local de $U(x)$), temos para energias próximas de $U(0)$ que o movimento da partícula é oscilatório. No caso em que $x=0$ é instável (máximo local de $U(x)$), caso o movimento se mantenha limitado -- se $U(x)$ forma uma barreira de potencial em ambos os lados do ponto de equilíbrio, por exemplo -- o movimento também será oscilatório, no sentido de que terá pontos de retorno em ambos os lados. Chamaremos esse tipo de sistema de \emph{oscilador} (unidimensional).

Em geral, vemos da equação (\ref{eq:eqmovimentoU}) que o movimento é anarmônico, isto é, que a trajetória da partícula não pode ser representada por uma função senoidal com frequência igual para todas as órbitas. Isso só acontece para $U(x) = k\,x^2/2$, em que a equação se reduz à do oscilador harmônico 
\begin{equation}\label{eq:eqmovimentoHarmonico}
m\,\ddot x + k\,x = 0\, ,
\end{equation}
cujo movimento tem frequência $\omega_0 = \sqrt{k/m\,}$ independente da amplitude (ou energia) da órbita.

\begin{figure}
    \centering
    \includegraphics[width= 0.99\columnwidth]{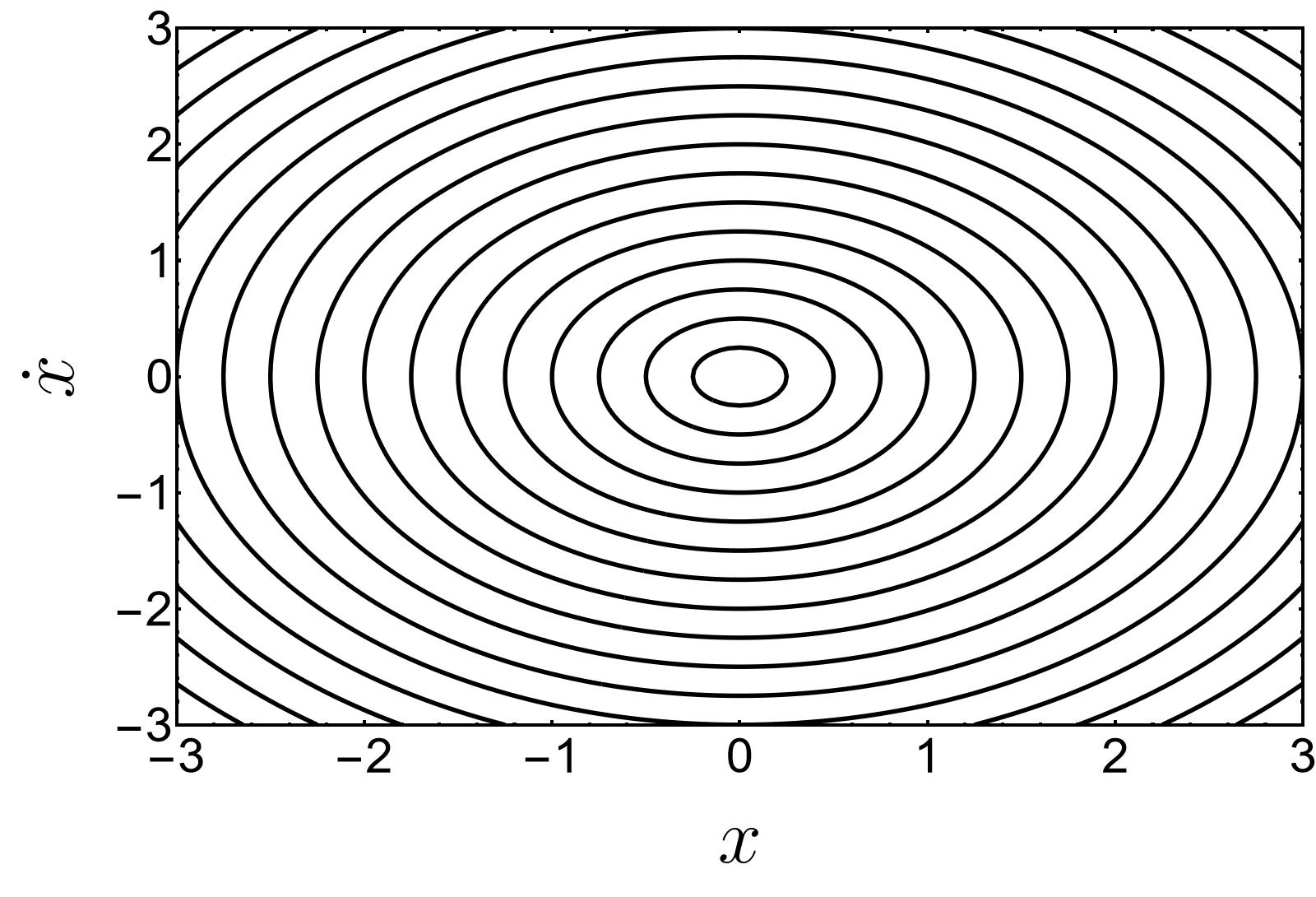}
    \caption{Diagrama de fases do oscilador harmônico, equações~(\ref{eq:eqmovimentoHarmonico}) e (\ref{eq:energiaHarmonico}), com $\omega_0 = \sqrt{k/m} =1$. Note que as escalas nos eixos horizontal e vertical são diferentes.}
    \label{fig:1}
\end{figure}

O diagrama de fases do oscilador harmônico está re\-pre\-sen\-ta\-do na figura \ref{fig:1}. O lugar geométrico das órbitas no espaço de fases é dado pela conservação de energia mecânica $E$ do sistema:
\begin{equation}\label{eq:energiaHarmonico}
E = \frac{1}{2}\,m\dot{x}^2 + \frac{1}{2}k\,x^2\, .
\end{equation}
Para cada valor distinto de $E>0$ a equação acima nos dá uma elipse simétrica em relação à origem, com semieixo horizontal $\sqrt{2E/k\,}$ e vertical $\sqrt{2E/m\,}$. As órbitas são então parametrizadas pela energia do sistema. Além disso, a razão entre os semieixos vertical e horizontal será $\omega_0 = \sqrt{k/m\,}$, a mesma para todas as órbitas. 

Nosso objetivo aqui é explorar, baseados em ferramentas úteis na análise do oscilador harmônico forçado, o que acontece quando o oscilador é anarmônico. Mas, primeiramente, vamos comentar sobre os osciladores não-lineares que serão nosso objeto de estudo.

\subsection{Oscilador quártico}

\begin{figure}
    \centering
    \includegraphics[width = 0.99\columnwidth]{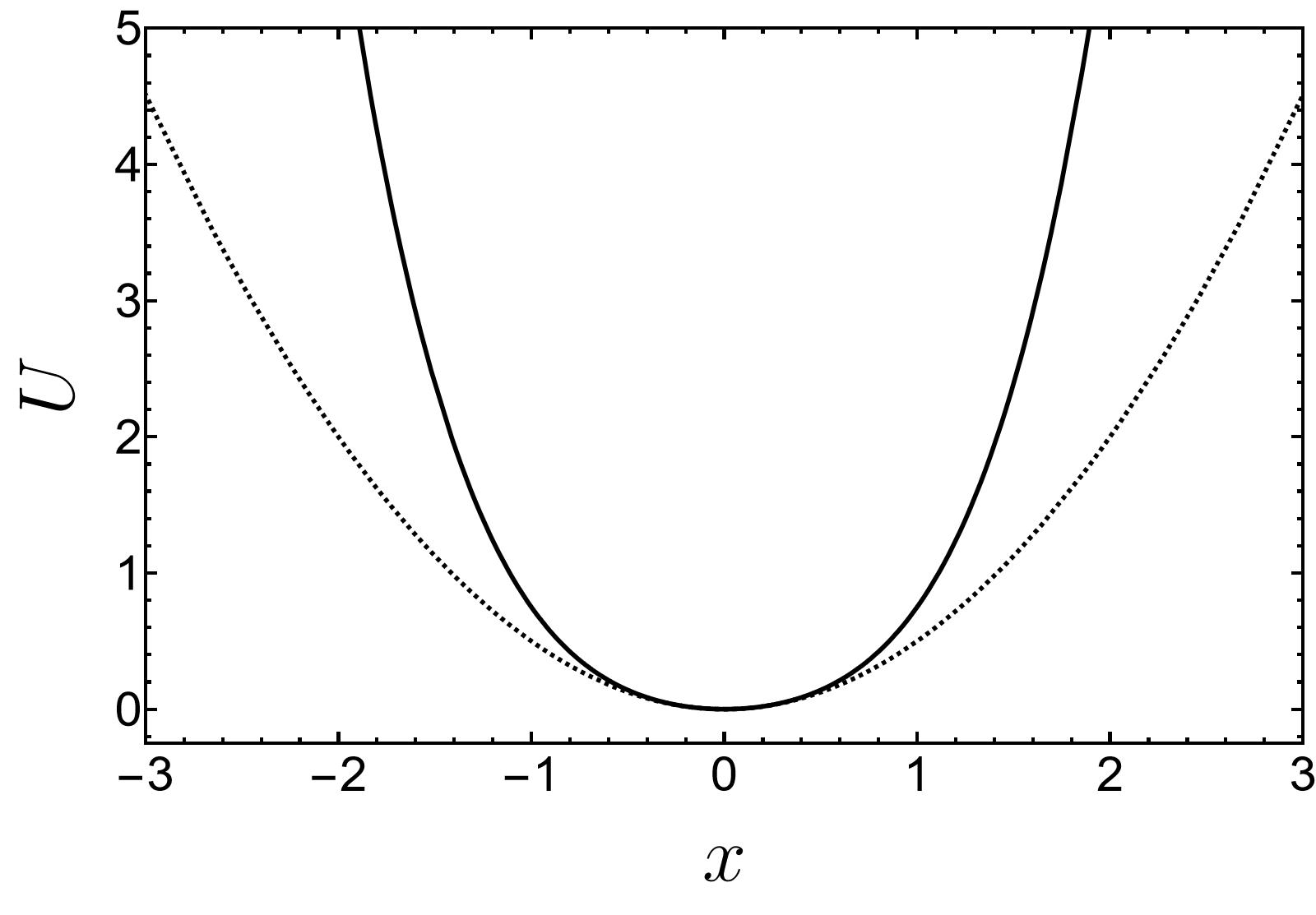}
    \caption{Energia potencial associada ao oscilador quártico (curva contínua), equação~(\ref{eq:energiapotencialQuartico}), com $k = 1$, $\ell = 1$ e à correspondente aproximação harmônica (curva pontilhada) com $k = 1$, $\ell = 0$.}
    \label{fig:2}
\end{figure}

Suponhamos que $x=0$ seja ponto de equilíbrio estável e que $U(x)$ seja par (simétrica com respeito ao ponto de equilíbrio) e crescente para $x>0$. 
Quando expandimos a energia potencial $U(x)$ em seu polinômio de Taylor ao redor de $x=0$ \cite{guidorizzi2001curso1} as potências ímpares de $x$ não aparecem, de modo que o termo seguinte ao quadrático é da ordem de $x^4$, levando a
\begin{equation}\label{eq:energiapotencialQuartico}
U(x) = \frac{1}{2}\,k\,x^2 + \frac{1}{4}\,\ell\,x^4\, ,
\end{equation}
com $k>0$, $\ell>0$. Isso nos leva à equação de movimento
\begin{equation}\label{eq:eqmovimentoQuartico}
m\,\ddot x + k\,x + \ell\,x^3 = 0\, , 
\end{equation}
na qual fica explícito o termo não-linear, proporcional a $x^3$.
O oscilador acima, chamado de \emph{oscilador quártico}, corresponde então à primeira correção ao movimento harmônico quando a amplitude das oscilações está fora do domínio da aproximação de ordem mais baixa na e\-ner\-gia potencial. Um gráfico da energia potencial de um oscilador quártico é apresentado na figura \ref{fig:2}, para os parâmetros $k=1$, $\ell=1$, juntamente com a cor\-res\-pon\-den\-te aproximação harmônica.

\begin{figure}
    \centering
    \includegraphics[width= 0.99\columnwidth]{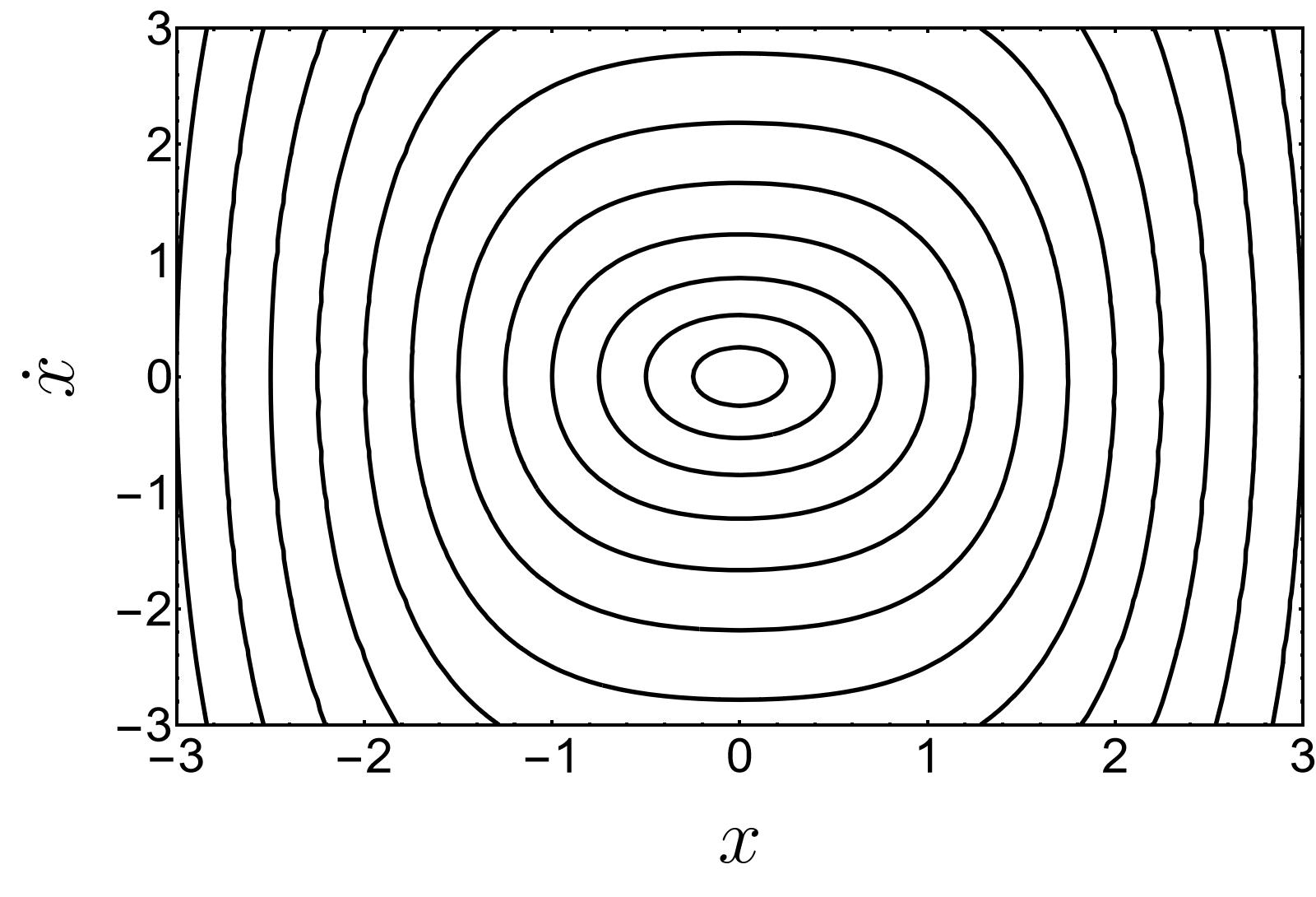}
    \caption{Diagrama de fases do oscilador quártico, equações~(\ref{eq:energiapotencialQuartico}) e (\ref{eq:eqmovimentoQuartico}), com $k/m = 1$, $\ell/m = 1$.}
    \label{fig:3}
\end{figure}

Como mencionado anteriormente, temos que no espaço de fases ($x, \dot{x}$) as órbitas de um oscilador harmônico são elipses perfeitas simétricas em relação aos eixos cartesianos, com razão entre os semieixos vertical e horizontal dada sempre por $\omega_0 = \sqrt{k/m}$ (figura \ref{fig:1}), que também é a frequência de oscilação. Vemos do diagrama de fases do oscilador quártico, figura \ref{fig:3}, que nele esse deixa de ser o caso. As elipses ficam ``distorcidas'', tanto mais quanto maior a amplitude da oscilação. Além disso, é possível mostrar que o período (\ref{eq:periodo}), e portanto a frequência de oscilação agora dependem da energia da órbita, e portanto de sua amplitude.

\subsection{Oscilador de Duffing}

\begin{figure}
    \centering
    \includegraphics[width = 0.99\columnwidth]{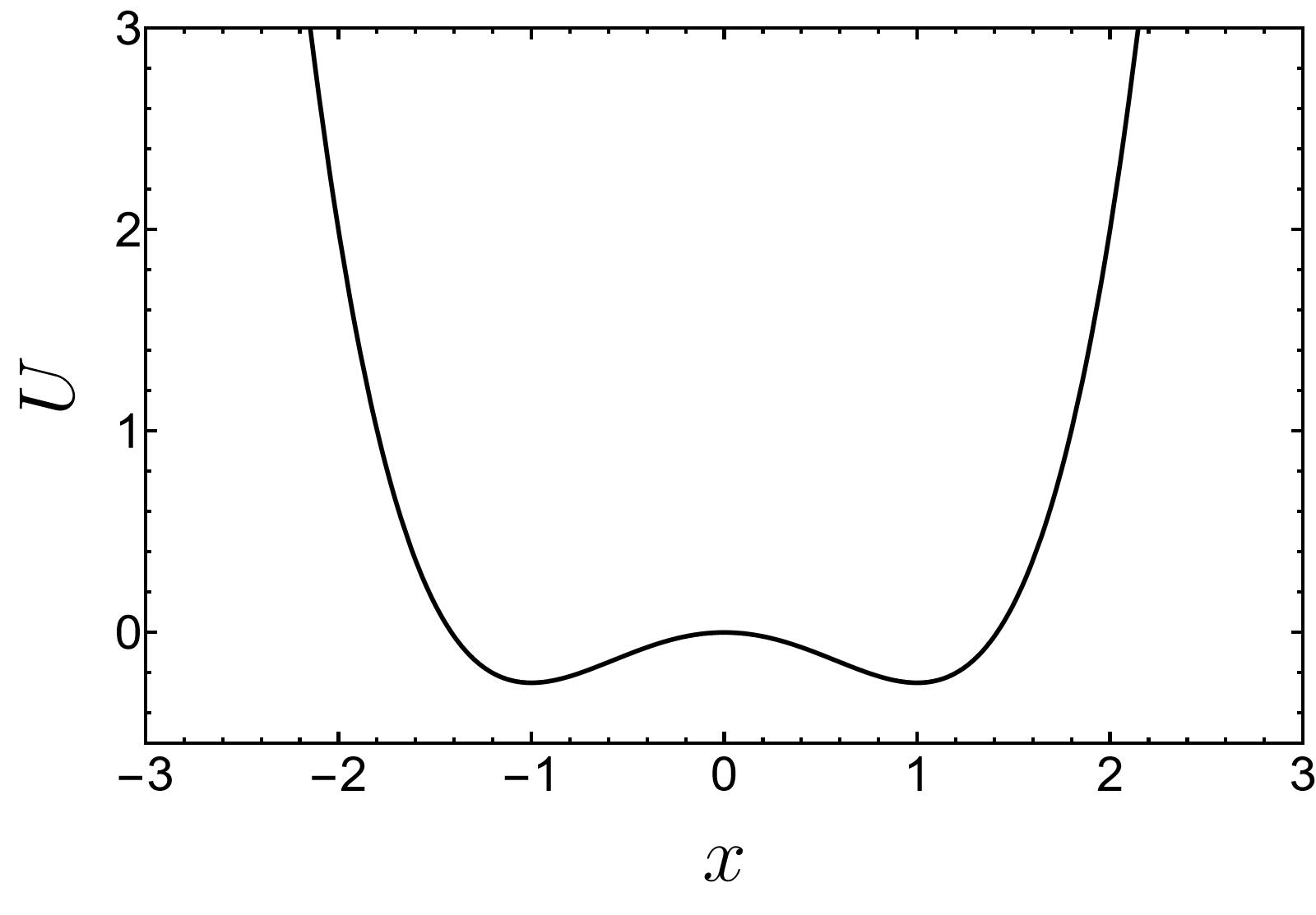}
    \caption{Energia potencial associada ao oscilador de Duffing com $k = 1$, $\ell = 1$.}
    \label{Potencial do Oscilador de Duffing}
\end{figure}

Um outro exemplo de oscilador corresponde a tomar a expansão na energia potencial até quarta ordem, mas considerar que o ponto de equilíbrio $x=0$ é instável. Temos nesse caso o chamado \emph{oscilador de Duffing}, com energia potencial 
\begin{equation}\label{eq:energiapotencialDuffing}
U(x) = - \frac{1}{2}\,k\,x^2 + \frac{1}{4}\,\ell\,x^4
\end{equation}
onde $k>0$, $\ell>0$. O gráfico dessa energia potencial é apresentado na figura \ref{Potencial do Oscilador de Duffing} para $k=1$, $\ell=1$ (compare com o gráfico para o oscilador quártico, figura~\ref{fig:2}). Vemos que, embora o ponto de equilíbrio central seja instável, o caráter oscilatório do movimento é garantido pelo termo quártico, que é dominante para valores altos de $|x|$, formando uma barreira de potencial já que $\ell>0$. 

\begin{figure}
    \centering
    \includegraphics[width = 0.99\columnwidth]{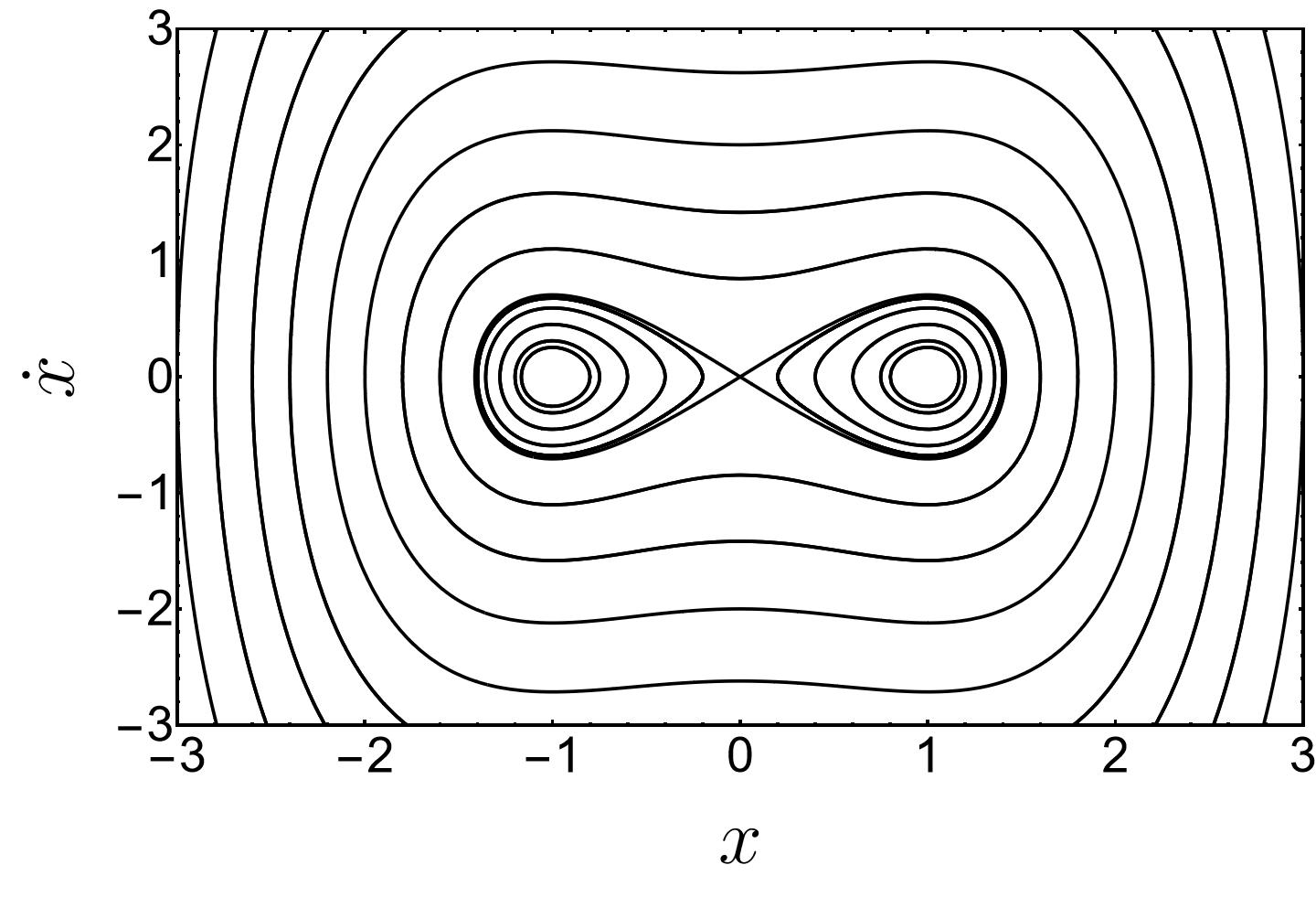}
    \caption{Diagrama de fases do oscilador de Duffing, equações~(\ref{eq:energiapotencialDuffing}) e (\ref{eq:eqmovimentoDuffing}), com $k/m = 1$, $\ell/m = 1$. É possível ver as órbitas em forma de ``infinito'' saindo e voltando para o ponto de equilíbrio instável, que correspondem a $E=0$.}   
    \label{fig:5}
\end{figure}

O diagrama de fases do oscilador de Duffing é apresentado na figura \ref{fig:5}, para os mesmos parâmetros utilizados na figura \ref{Potencial do Oscilador de Duffing}. Vemos que o ponto de equilíbrio instável em $x=0$ tem, associado à sua mesma energia mecânica $E=0$, duas outras órbitas: uma para $x>0$ e outra para $x<0$ (que juntas formam um símbolo parecido com o de `infinito'). Essas órbitas, chamadas de \emph{separatrizes}, são divisoras entre o movimento que ocorre só em um dos lados e o movimento que ocorre tendo ambos os lados disponíveis.

A equação de movimento para o oscilador de Duffing é então
\begin{equation}\label{eq:eqmovimentoDuffing}
m\,\ddot x - k\,x + \ell\,x^3 = 0 \, ,
\end{equation}
novamente mostrando uma característica não-linear.

\section{Osciladores forçados, mapas estroboscópicos e caos}

\begin{figure}
    \centering
    \includegraphics[width = 0.99\columnwidth]{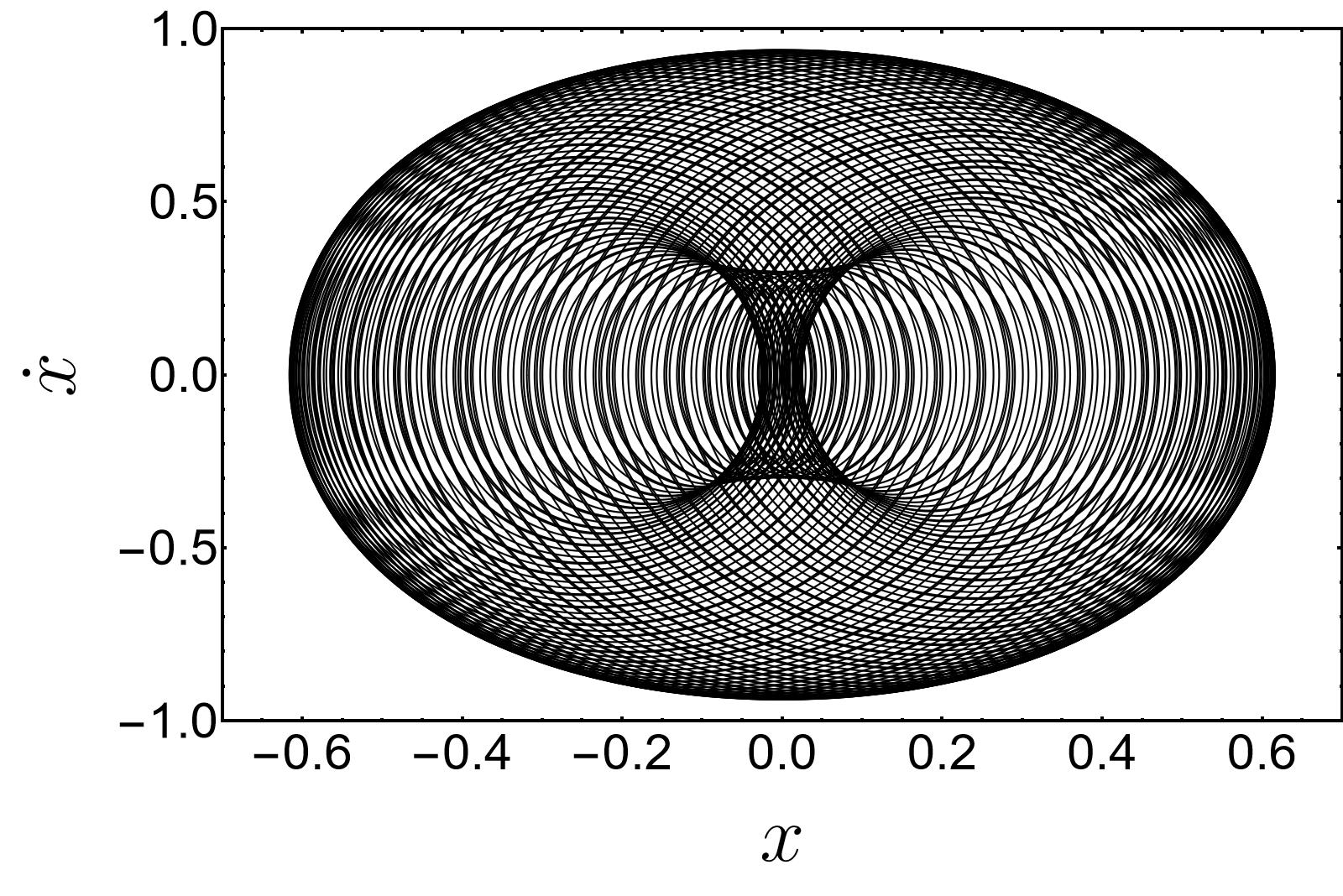}
    \caption{Uma órbita do oscilador harmônico perturbado, equação~(\ref{eq:eqmovimentoOHforcado}), com $\omega_0=1$, $F_0/m = 1$, $\omega = 2 \pi/3$  e condições iniciais $x_{0} = 0$, $\dot{x}_{0} = 0.125$.}
    \label{fig:6}
\end{figure}

Consideremos uma força externa periódica no tempo aplicada ao oscilador, da forma
\begin{equation}\label{eq:F}
F(t) = F_0 \cos(\omega t)\,. 
\end{equation}
A equação de movimento para o oscilador passa a ser uma equação não homogênea, dependente do tempo:
\begin{equation}
m\,\ddot x + \frac{dU}{dx} = F_0 \cos(\omega t)\,.
\end{equation}
Assim não há mais conservação de energia mecânica, e o diagrama de fases não é mais `folheado' por órbitas com energias distintas.
Mesmo para o caso do oscilador harmônico, que representa a primeira aproximação para pequenas oscilações ao redor do ponto de equilíbrio estável $x=0$, cuja equação de movimento se reduz a 
\begin{equation}\label{eq:eqmovimentoOHforcado}
\,\ddot x + \omega_0^2\, x = \frac{F_0}{m} \cos(\omega t)\,,
\end{equation}
com $\omega_0 = \sqrt{k/m\,}$ sua frequência natural de oscilação,
uma única órbita genérica é capaz de percorrer toda uma região  do espaço de fases de maneira que seria impossível ter uma descrição qualitativa, visual do movimento (figura \ref{fig:6}). Apesar disso, a solução do problema pode ser obtida analiticamente e é bem conhecida, onde consideramos $\omega\neq\omega_0$ \cite{watarimecanicaV1}:
\begin{equation}\label{eq:solucaoOHF}
    x(t) = C_{1} \cos(\omega_0 t + \phi) + \frac{F_0}{m} \,\frac{\cos(\omega t)}{{\omega_{0}}^2 - \omega^2}\,.
\end{equation}
Desse modo, seria interessante termos algum método que recobrasse, no espaço de fases, propriedades intrínsecas do oscilador harmônico, como por exemplo alguma relação entre as órbitas na presença de $F(t)$ e as correspondentes soluções do caso não perturbado (sem força externa).

\begin{figure}
    \centering
    \includegraphics[width = 0.99\columnwidth]{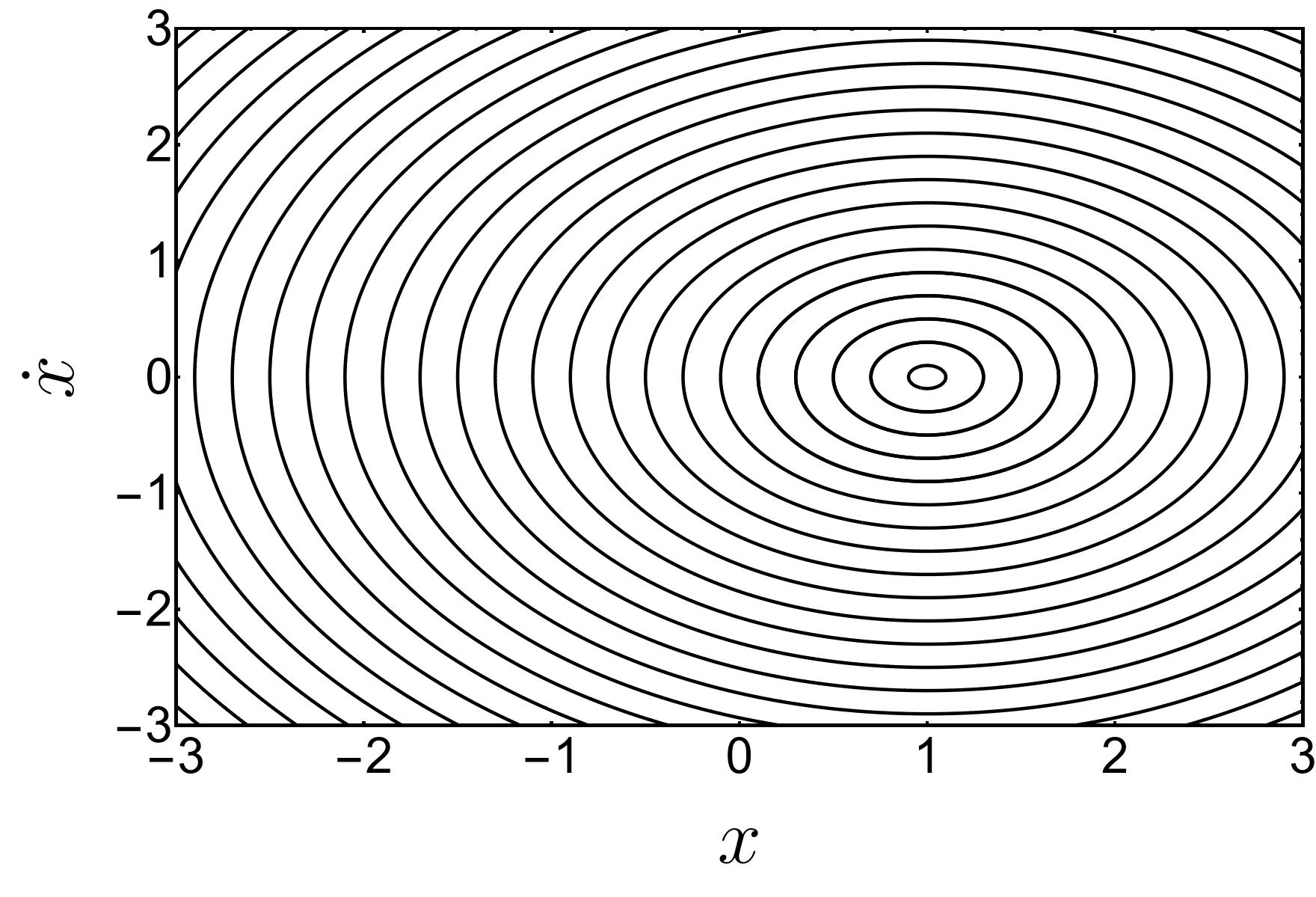}
    \caption{Diagrama de fases do oscilador harmônico sujeito a uma força externa constante $F_0$, com parâmetros $\omega_0=1$, $F_0/m=1$, $\omega = 0$. As curvas são elipses da forma (\ref{eq:elipsesOHF0}), parametrizadas pela amplitude $C_1$.}
    \label{fig:7}
\end{figure}

Consideraremos no que segue o termo de força externa como uma perturbação, referindo-nos a esse caso como um `oscilador forçado' ou `perturbado' e, no caso de ausência de força externa, como o `oscilador não perturbado'.
Primeiramente, vejamos como fica o diagrama de fases quando a força é constante (o que e\-qui\-va\-le ma\-te\-ma\-ti\-ca\-men\-te a considerarmos $\omega=0$ na equação~(\ref{eq:eqmovimentoOHforcado})). Nesse caso, a equação admite uma solução particular constante $x_{\rm p} = F_0/(m\omega_0^2)$, e portanto a solução geral é da forma~(\ref{eq:solucaoOHF}) com $\omega=0$. As órbitas no diagrama de fases então continuam sendo elipses com razão $\omega_0$ entre os semieixos, mas deslocadas da origem no eixo horizontal por $a_0 = F_0/(m\omega_0^2)$ (ver figura~\ref{fig:7}):
\begin{equation}\label{eq:elipsesOHF0}
    \frac{(x - a_0)^2}{C_1^2} + \frac{\dot{x} ^2}{\ C_1^2\,\omega_0^2\ } = 1\,.
\end{equation}
Dessa forma, como na presença de uma força externa cons\-tan\-te o sistema continua autônomo (independente do tempo), possuindo um diagrama de fases bem definido (figura \ref{fig:7}), o método procurado para analisar a força dependente do tempo deveria recobrar também características dessas elipses deslocadas.

É possível mostrar por argumentos de sistemas dinâmicos \cite{deaguiarLivro, tabor1989chaos} que essas características do oscilador harmônico aparecem no caso forçado (\ref{eq:eqmovimentoOHforcado}) quando con\-si\-de\-ra\-mos, no espaço de fases $(x, \dot{x})$, \emph{mapas estroboscópicos} (ou \emph{seções estroboscópicas}) do movimento com um período determinado. A cada período $T=2\pi/\omega$ da força externa, isto é, para cada 
\begin{equation}
    t_n = \frac{2n\pi}{\omega}\,,
\end{equation} 
plotamos no espaço de fases um ponto representando a posição e velocidade da partícula naquele instante. Isso pode ser visto como a sobreposição de um conjunto de `fotos' ou \emph{`flashes'} uniformemente espaçados no tempo, com frequência $\omega$ igual à frequência da força externa (daí o termo `estroboscópico'). O conjunto de pontos plotados no plano $x-\dot{x}$ segundo esse mapa, a partir de uma dada condição inicial, chamaremos de \emph{órbita do mapa estroboscópico} com essas condições iniciais e, quando não houver perigo de confusão, utilizaremos o termo `mapa estroboscópico' tanto para o processo de plotar os pontos quanto para o conjunto de órbitas que esse mapa produz no espaço de fases, correspondência que ficará evidente adiante. Da solução geral das equações de movimento é possível então mostrar que esses pontos cairão sempre sobre uma elipse no espaço de fases, cuja razão dos semieixos será sempre igual a $\omega_0$ (como no caso do diagrama de fases do oscilador harmônico não perturbado, ver figura~\ref{fig:1}). Seu centro, no entanto, fica deslocado da origem, assemelhando-se ao caso de força externa constante (figura~\ref{fig:7}). 
De fato, calculando $\dot{x}(t)$ da equação~(\ref{eq:solucaoOHF}) e depois fazendo $t_n = 2 n \pi/\omega$, temos que 
\begin{equation}\label{eq:elipsedeslocadaomega}
    \frac{(x_n - a)^2}{C_1^2} + \frac{ (\dot{x}_n) ^2}{\ C_1^2\,\omega_0^2\ } = 1\,,
\end{equation}
onde $a=F_0/[m ({\omega_{0}}^2 - \omega^2)]$ e $x_n = x(t_n)$, $\dot{x}_n = \dot{x}(t_n)$.

Assim, vemos que mapas estroboscópicos com frequência igual à da força externa nos permitem visualizar no espaço de fases características da órbita forçada que ficariam camufladas se olhássemos para a órbita `completa', contínua. É esperado então que essa ferramenta seja útil na análise do movimento de sistemas mais complicados, como por exemplo os osciladores a\-nar\-mô\-ni\-cos forçados.



\begin{figure}
    \centering
    \includegraphics[width = 0.99\columnwidth]{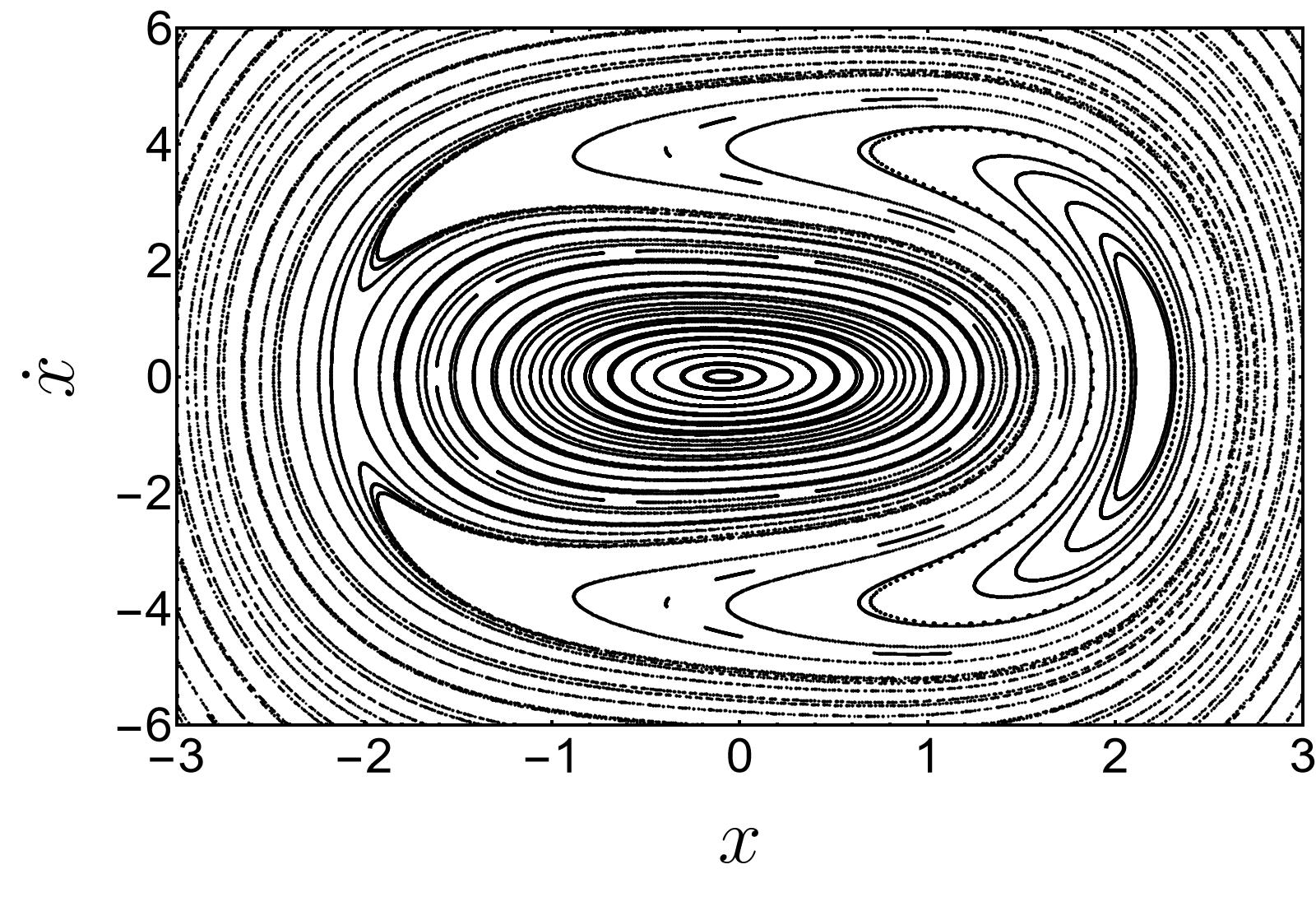}
    \caption{Órbitas do mapa estroboscópico para o oscilador quártico perturbado, com $k=1$, $\ell=1$, $m=1$, $F_0 = 0.3$, $\omega = 2 \pi/3$. É possível ver, além das elipses distorcidas, a formação de uma ilha de estabilidade distorcida na região à direita. Note que, como $\omega>\sqrt{k/m}$, teríamos $a<0$ na equação~(\ref{eq:elipsedeslocadaomega}) para a aproximação harmônica, de maneira que as elipses distorcidas ``centrais'' da figura ficam deslocadas para a esquerda de $x=0$.}
    \label{fig:8}
\end{figure}

\subsection{Oscilador quártico forçado}

Como já mencionado, o próximo termo na expansão em série de potências da energia potencial $U(x)$ ao redor do ponto de equilíbrio estável $x=0$, se mantivermos simetria de reflexão, é um termo quártico. Ele deve aparecer para movimentos com energias mais altas (e portanto para amplitudes maiores), nas quais o termo de segunda ordem não é suficiente para descrever a oscilação. 

Diferentemente do oscilador harmônico, em que a equação de movimento é linear e portanto é possível obter a solução geral como uma combinação linear de soluções fundamentais da equação homogênea mais uma solução particular da equação não-homogênea \cite{boyce2010equaccoes}, a não-linearidade da equação de movimento (\ref{eq:eqmovimentoQuartico}) não permite escrever sua solução geral como uma soma de soluções fundamentais, nem somar uma solução da equação homogênea com uma solução particular envolvendo a força externa (\ref{eq:F}). Assim, embora seja possível resolver por quadraturas a equação homogênea, a equação com termo de força externa precisa ser resolvida separadamente para cada par de condições iniciais $(x_0, p_0)$ diferentes. Além disso, em geral, não é possível obter expressão analítica na presença do termo de força dependente do tempo (nem por quadraturas), de modo que temos que recorrer a soluções numéricas.

Mesmo com essas dificuldades, é possível obter numericamente os mapas estroboscópicos no espaço de fases, como no caso do oscilador harmônico. Baseando-nos na discussão feita acima, é esperado que esses mapas `reflitam' algumas características do oscilador não perturbado, em particular a distorção das elipses.

\begin{figure}
    \centering
    \includegraphics[width = 0.99\columnwidth]{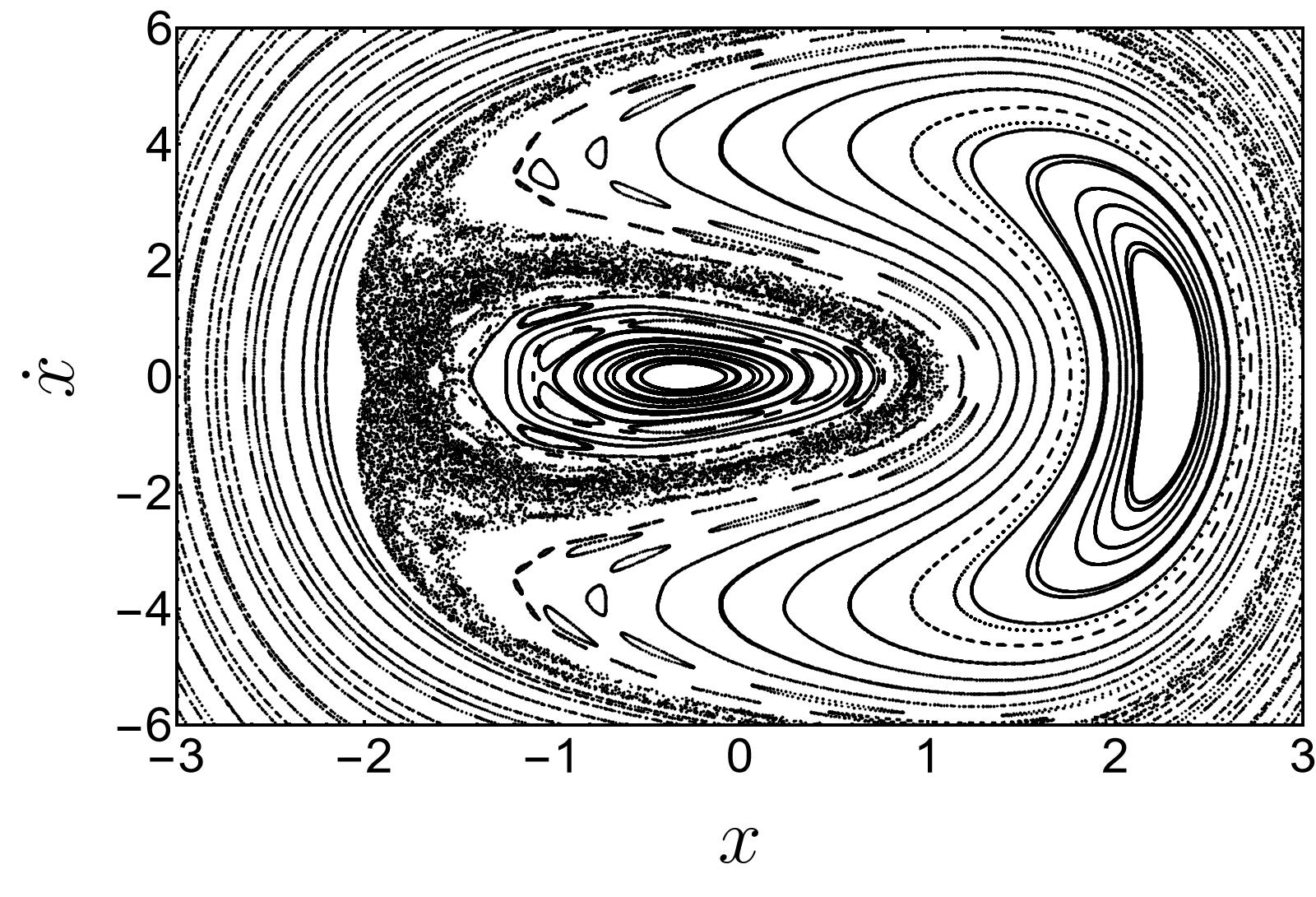}
    \caption{Órbitas do mapa estroboscópico para o oscilador quártico perturbado, com $k=1$, $\ell=1$, $m=1$, $F_0 = 1$, $\omega = 2 \pi/3$. É possível ver uma região caótica sendo formada ao redor da ilha central de estabilidade (deslocada para a esquerda, ver figura~\ref{fig:8}), assim como outras ilhotas tanto na região central quanto na região estável à direita.}
    \label{fig:9}
\end{figure}

De fato, para $F_0/m$ pequeno, os mapas estroboscópicos mostram elipses distorcidas (ver figura \ref{fig:8}), como no caso do diagrama de fases não perturbado (figura \ref{fig:3}), assim como também aparecem outras estruturas à di\-rei\-ta, órbitas formando ``ilhas'' distorcidas. No entanto, ao aumentarmos a amplitude $F_0/m$ da força externa, um fenômeno novo ocorre: além da esperada distorção das órbitas conforme a força aumenta, começam a aparecer regiões no espaço de fases em que uma órbita sozinha preenche uma área no plano $x-\dot{x}$ pela aplicação do mapa estroboscópico às suas condições iniciais, não estando mais restrita a uma curva (figura~\ref{fig:9}). 
Esse fenômeno não acontece no oscilador harmônico, e está intrinsecamente ligado à não-linearidade do oscilador (neste caso o termo cúbico na equação de movimento, ou o termo quártico na energia potencial), quando há a presença da força externa.

Temos então algo inicialmente inesperado: a técnica dos mapas estroboscópicos parece não reproduzir as curvas distorcidas para amplitudes grandes da força externa. Surge a pergunta: isso ocorre por uma limitação do método, ou há algum fenômeno intrínseco a esse tipo de sistema, que apareceria em qualquer outro método de análise? A resposta a essa pergunta exige co\-nhe\-ci\-men\-tos mais aprofundados da dinâmica de sistemas hamiltonianos, e pretendemos dar uma breve introdução a esses conceitos na última seção deste artigo (para uma exposição de\-ta\-lha\-da do assunto, ver as referências \cite{deaguiar1994RBEF, tabor1989chaos, lichtenbergLieberman1992, deaguiarLivro}). Mas já adiantamos aqui a resposta. Fenômenos como esse são intrínsecos a sistemas não-lineares dependentes do tempo ou multidimensionais e são uma assinatura da dinâmica caótica nas regiões correspondentes do espaço de fases. 

\begin{figure*}
    \centering
    
    \includegraphics[width = 0.99\columnwidth]{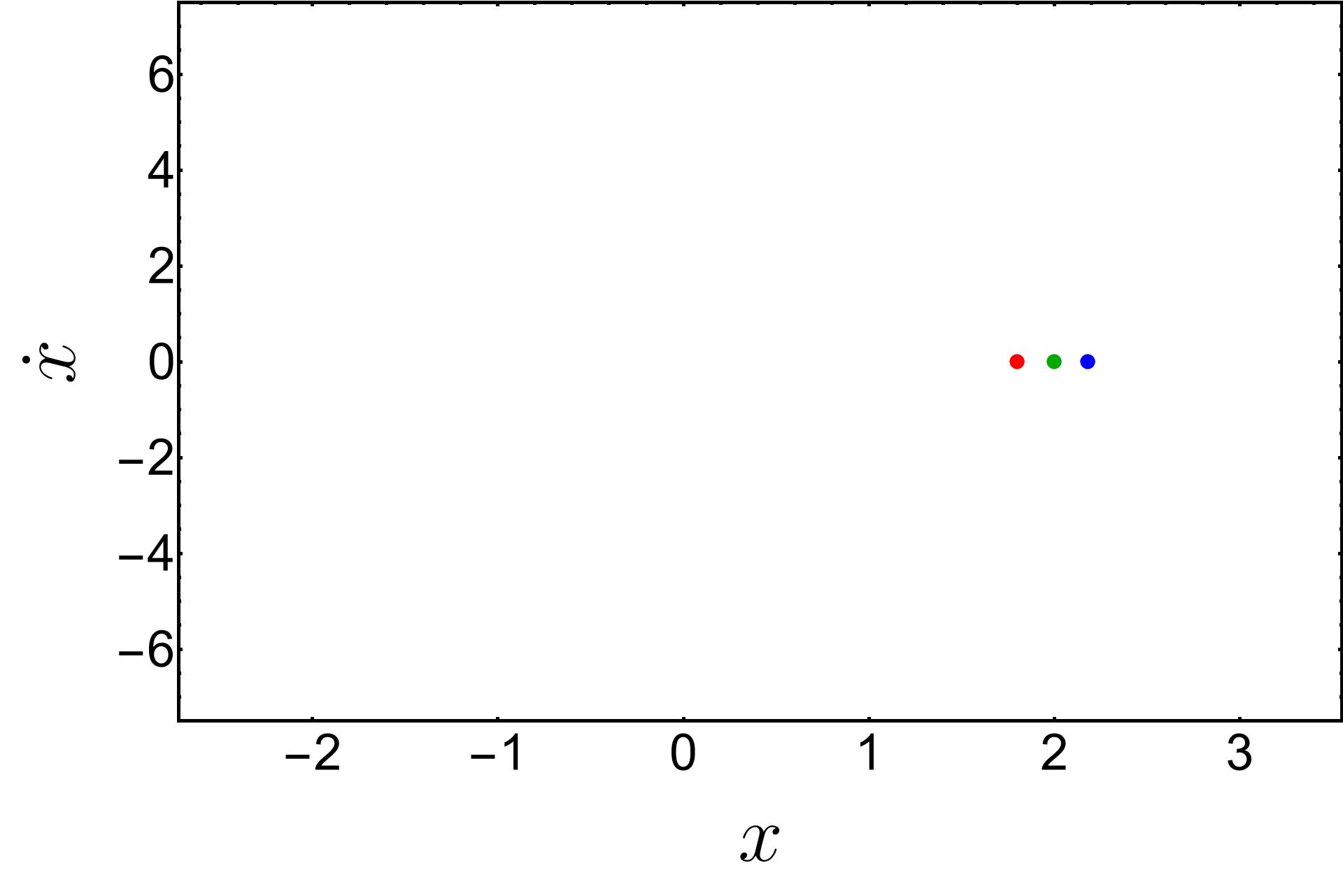} \qquad
    \includegraphics[width = 0.99\columnwidth]{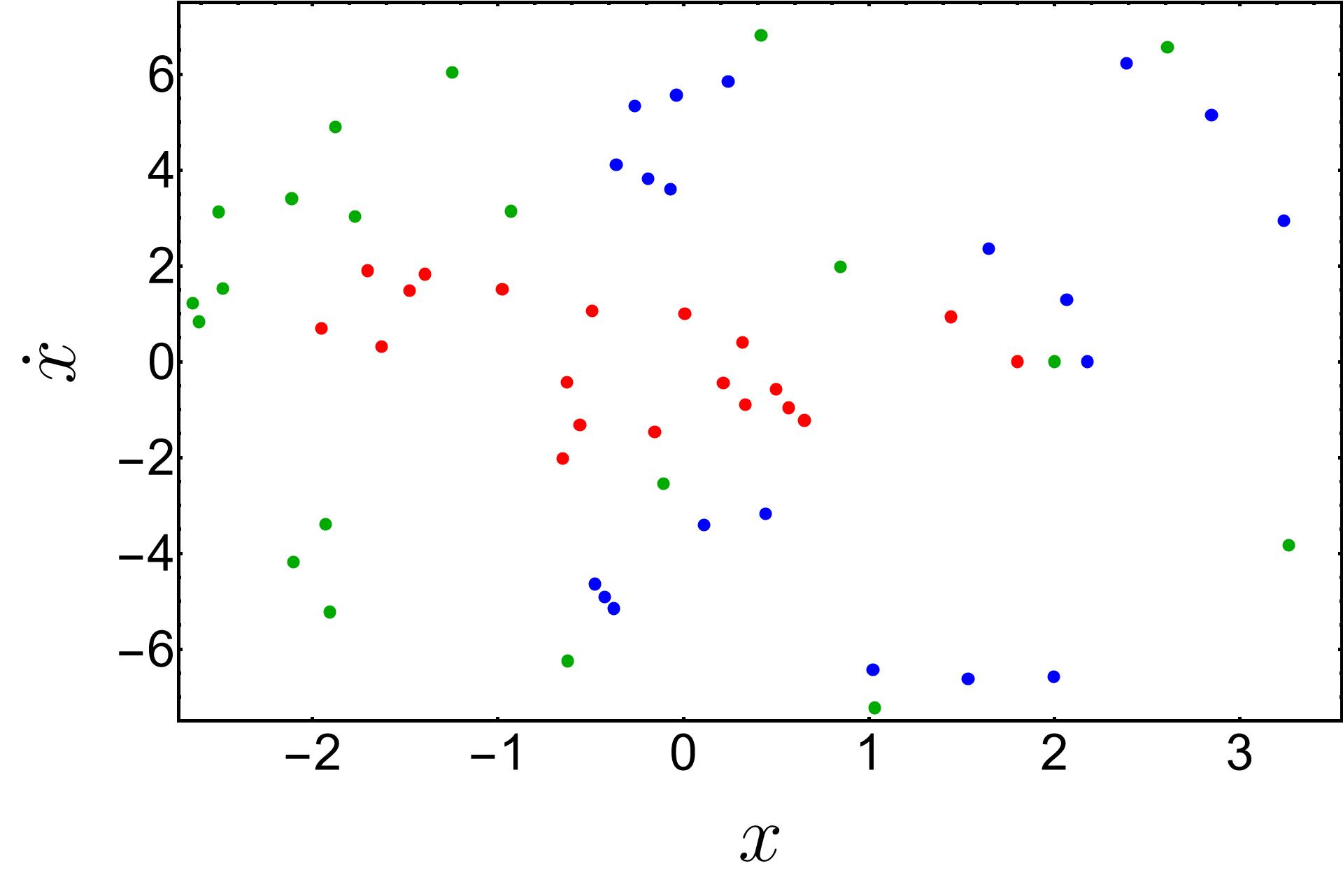}
    \\ \hfill \\
    \includegraphics[width = 0.99\columnwidth]{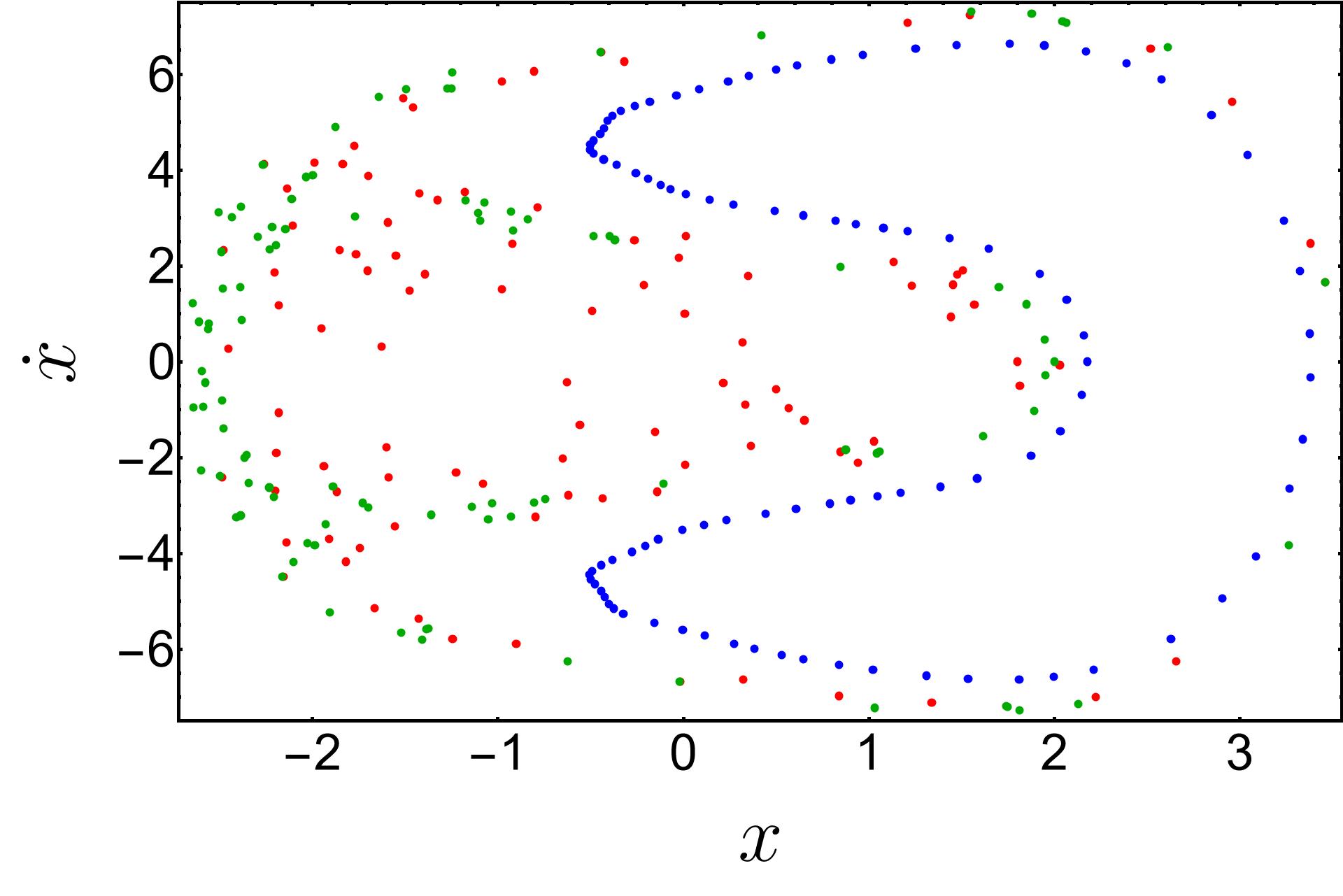} \qquad
    \includegraphics[width = 0.99\columnwidth]{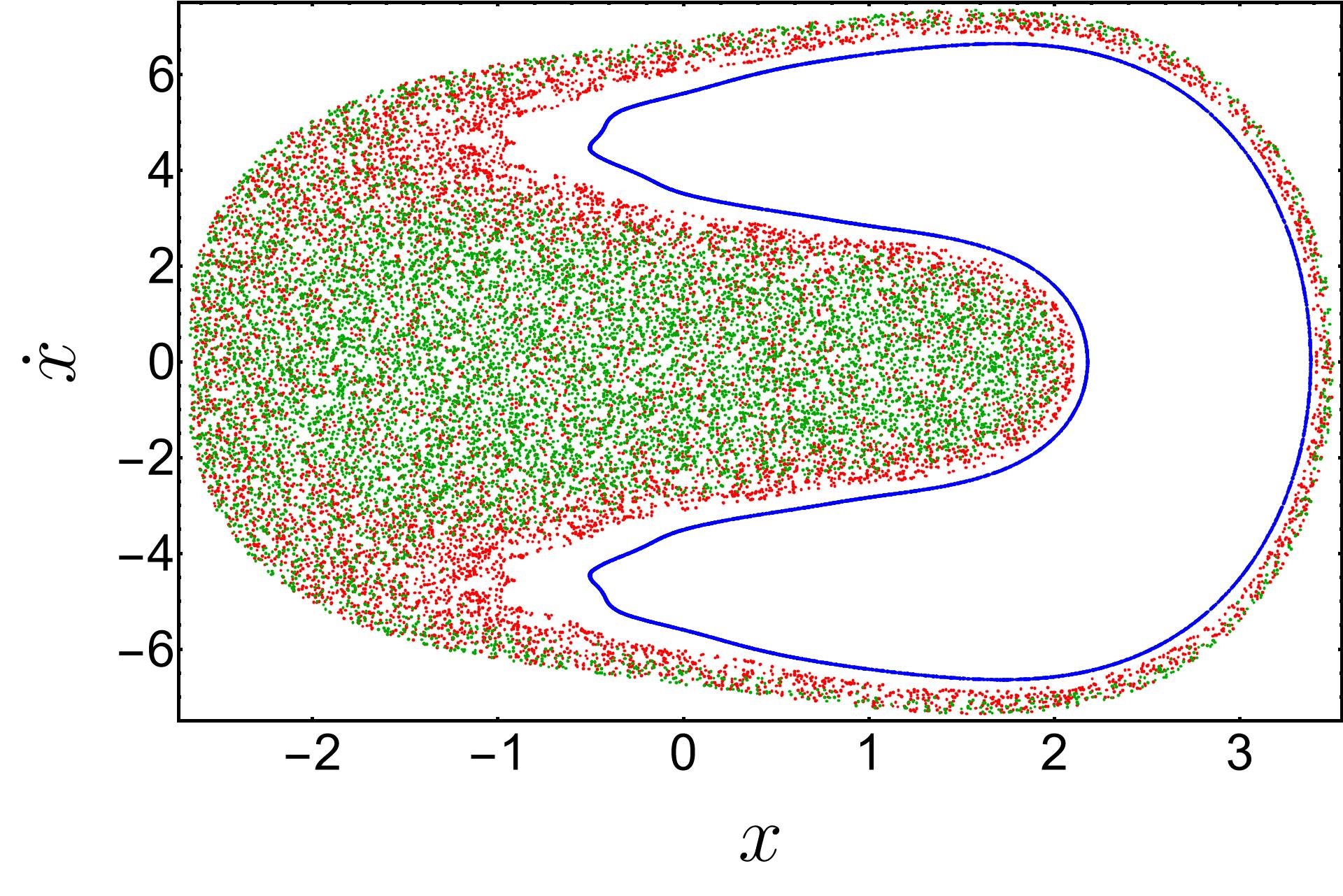}
    \caption{Quadro representando a evolução temporal de condições iniciais próximas, em vermelho $(x(0),\dot{x}(0)) = (1.8,0)$, em verde $(x(0),\dot{x}(0)) = (2,0)$ e em azul $(x(0),\dot{x}(0)) = (2.18,0)$, no oscilador de Duffing perturbado com $k=1$, $\ell=1$, $m=1$, $F_0 = 1$,  $\omega = 2 \pi/3$. \emph{Acima, esquerda:} condições iniciais do movimento para cada órbita. \emph{Acima, direita:} correspondente mapa estroboscópico dessas condições iniciais para os 19 primeiros períodos (20 pontos para cada órbita). \emph{Abaixo, esquerda:} mapa estroboscópico para os 100 primeiros pontos. \emph{Abaixo, direita:} mapa estroboscópico para os 10000 primeiros pontos. Vemos que, no início do movimento, não fica clara a estrutura de cada órbita na seção. Conforme o tempo passa e os pontos vão sendo plotados, começamos a enxergar uma estrutura unidimensional, como no caso da órbita regular azul, ou a falta dessa estrutura resultando no preenchimento de toda uma área, como no caso das órbitas caóticas vermelha e verde; além disso, vemos que as órbitas vermelha e verde vão gradativamente preenchendo praticamente a mesma área (sem nunca se cruzar), característica da imprevisibilidade associada à incerteza nas condições iniciais.
    }
    \label{fig:10}
\end{figure*}

Mas o que significa uma região ser caótica? In\-tui\-ti\-va\-men\-te, no contexto que estamos analisando, vemos que as regiões denominadas caóticas, em que órbitas não preenchem curvas no plano $x-\dot{x}$ do mapa estroboscópico mas sim toda uma região bidimensional, são tais que dadas duas condições iniciais genéricas muito próximas nessa região as órbitas correspondentes não estarão fixas sobre duas curvas geradas pelo mapa de modo a podermos compará-las ao longo do tempo. Na verdade, ambas preencherão a área toda dessa região, e para tempos muito longos fica impossível na prática comparar suas posições e velocidades no mesmo instante de tempo. Em outras palavras, se houver uma incerteza nas condições iniciais do movimento, por menor que seja, o estado final do sistema não poderá ser predito para tempos longos; a incerteza se propagará exponencialmente ao longo do tempo.
Ilustramos na figura~\ref{fig:10} o processo iterativo de plotar os pontos segundo o mapa estroboscópico, para três condições iniciais, uma correspondente a uma órbita regular (azul) e outras duas a órbitas caóticas (vermelha e verde).\footnote{Esse processo é ilustrado na figura~\ref{fig:10} para o oscilador de Duffing forçado, que será discutido na subseção seguinte; contudo, para fins ilustrativos, não há problema em apresentar essa figura agora.}
Vemos então que com o tempo os pontos correspondentes à órbita azul preenchem uma curva, unidimensional, recebendo portanto a qua\-li\-fi\-ca\-ção de ``regular'' por comparação com o caso não perturbado. Já a órbita correspondendo à condição inicial vermelha preenche densamente, sozinha, uma área (região bidimensional) no espaço de fases, o que reflete a sensibilidade a condições iniciais (justificando o nome ``caótica''); a outra órbita (verde) com condição inicial muito próxima, dentro dessa área vermelha preenchida, também preencherá (densamente) praticamente toda a região bidimensional, de maneira que uma incerteza nas condições iniciais não permitiria acompanhar continuamente um ``tubo'' de órbitas.

Assim, órbitas regulares estão associadas a regiões preenchidas por curvas segundo o mapa estroboscópico, enquanto que órbitas caóticas estão associadas a regiões em que os mapas estroboscópicos não apresentam estrutura definida no espaço de fases (no sentido discutido acima), com as órbitas do mapa preenchendo densamente, cada uma, uma região bidimensional no plano.
Sabemos que esse comportamento é devido a ressonâncias entre as frequências de oscilação~do sistema (a frequência do movimento não perturbado e a frequência da força externa \cite{deaguiarLivro,lichtenbergLieberman1992}); isso não aparece apenas no oscilador quártico, mas também em outros osciladores não-lineares e no pêndulo forçado (cuja equação de movimento também é não-linear, como discutiremos posteriormente).

\begin{figure}

    \includegraphics[width = 0.99 \columnwidth]{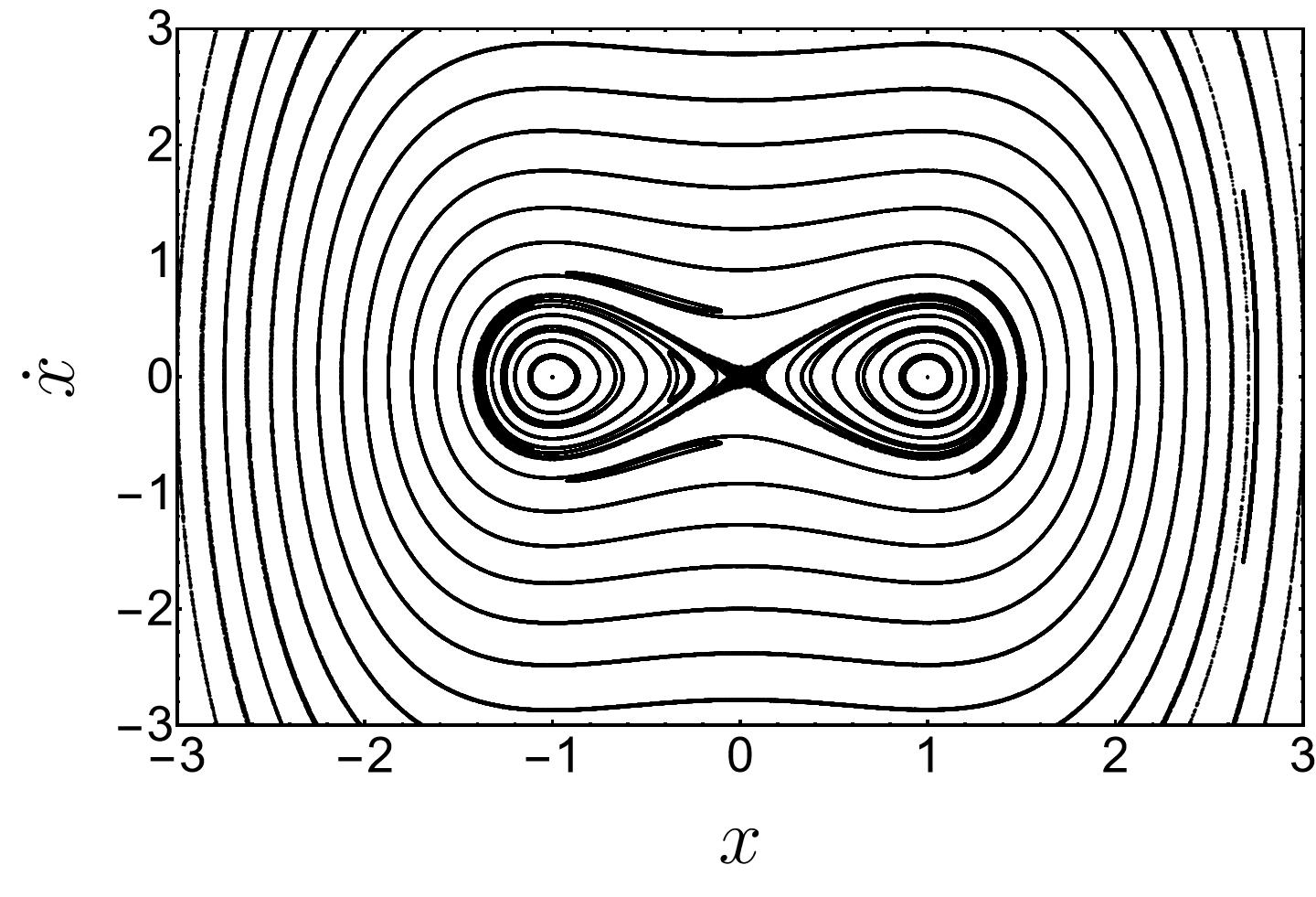}
    \\ \hfill \\
    \includegraphics[width = 0.99\columnwidth]{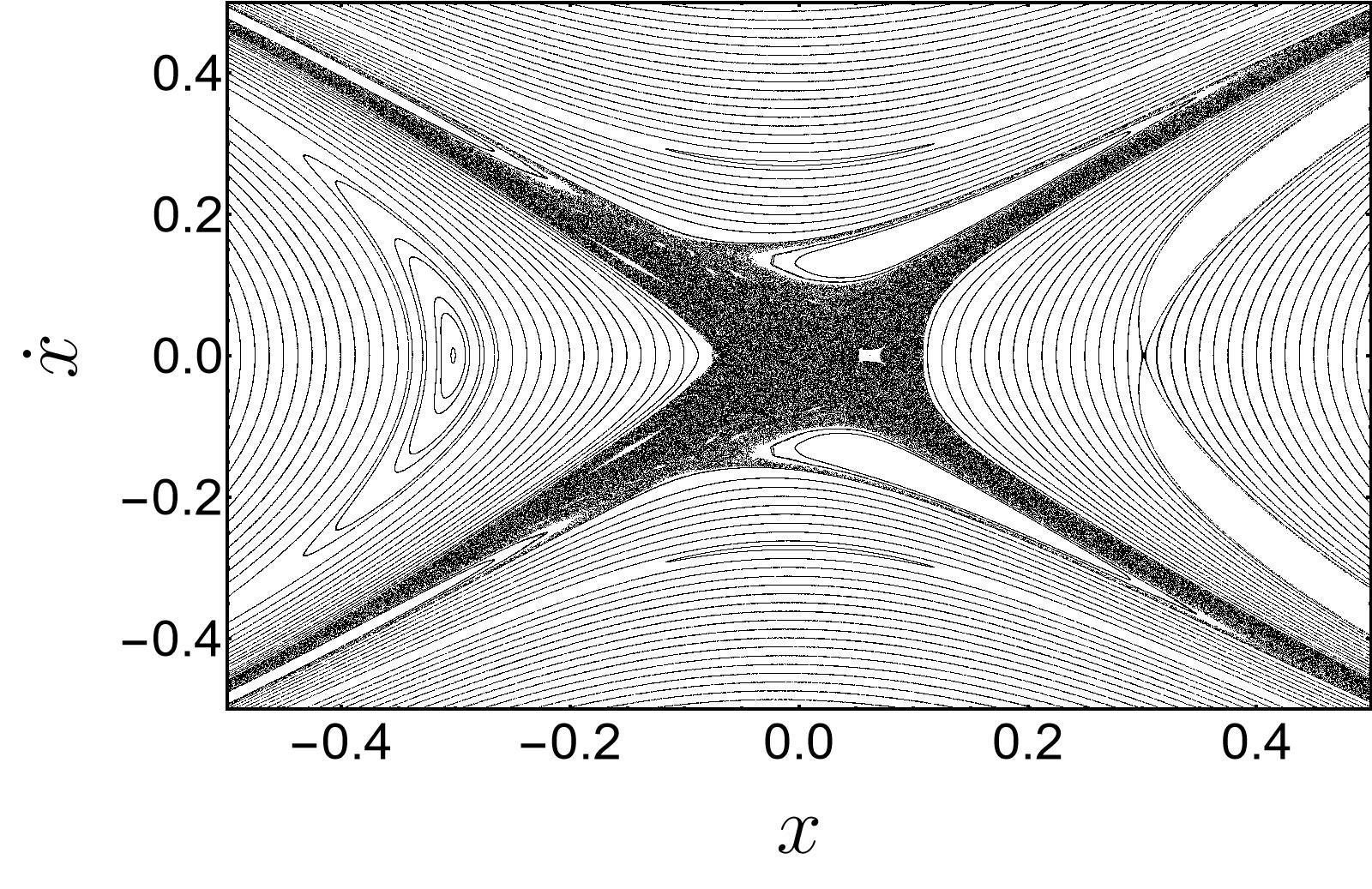}

    \caption{\emph{Acima:} Órbitas do mapa estroboscópico para o oscilador de Duffing perturbado, com $k =1$, $\ell=1$, $m=1$, $F_0 = 0.003$, $\omega = 2 \pi/3$. É possível ver uma região caótica próximo da separatriz e algumas ilhas de estabilidade na parte esquerda. \emph{Abaixo:} zoom da região central, mostrando de\-ta\-lhes da quebra da separatriz em uma região caótica.}
    \label{fig:11}
\end{figure}

\subsection{Oscilador de Duffing forçado}

No caso do oscilador de Duffing, quando o termo de força (\ref{eq:F}) é inserido na equação~(\ref{eq:eqmovimentoDuffing}), pode-se imaginar que condições iniciais próximas do ponto de equilíbrio instável ($x=0, \dot{x} = 0$) sejam levadas rapidamente para longe devido à instabilidade do movimento na região. Assim, intuitivamente, é esperado que regiões caóticas apareçam primeiro (ao se aumentar gradualmente a amplitude $F_0$ da força) próximo ao ponto de equilíbrio instável no espaço de fases do que em outras regiões (próximo dos pontos de equilíbrio estáveis com $x\neq 0$, por exemplo; ver figura~\ref{Potencial do Oscilador de Duffing}). 

\begin{figure}
    \centering
    \includegraphics[width = 0.99\columnwidth]{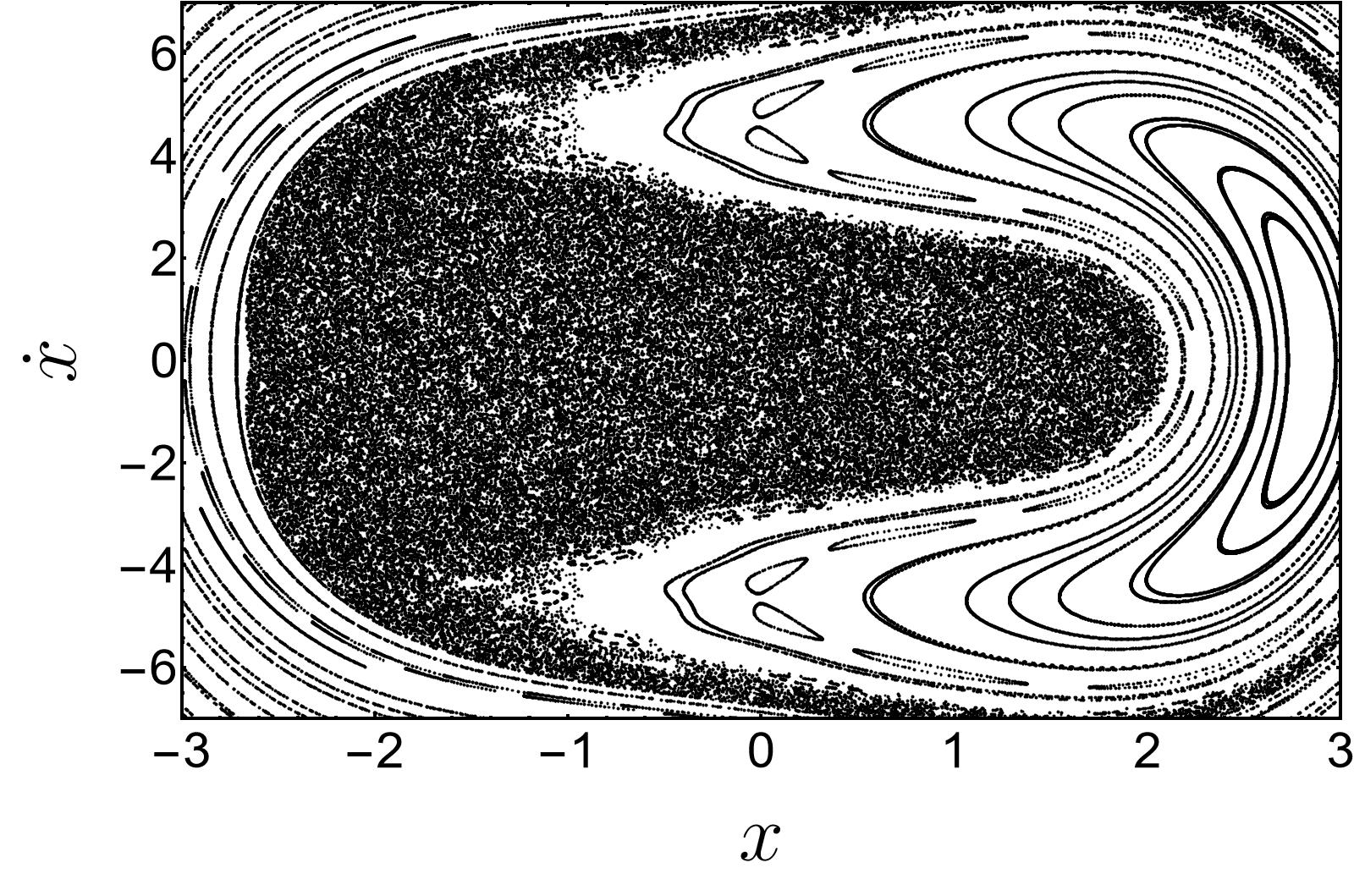}
    \caption{Órbitas do mapa estroboscópico para o oscilador de Duffing perturbado, com $k=1$, $\ell=1$, $m=1$, $F_0 = 1$, $\omega = 2 \pi/3$. É possível ver, na região central, a formação de um ``mar caótico'', devido à existência do ponto de equilíbrio instável na origem (comparar com o mapa para o oscilador quártico com os mesmos parâmetros, figura \ref{fig:9}).}
    \label{fig:12}
\end{figure}

E, de fato, é isso o que se vê. Diferentemente do caso do oscilador quártico, no oscilador de Duffing mesmo uma pequena amplitude da força externa já gera uma região considerável de caos próxima da origem do espaço de fases quando aplicamos o mapa estroboscópico às condições iniciais do plano, como exemplificado na figura \ref{fig:11}. Para amplitudes maiores $F_0$ da perturbação, o que era antes uma região central regular, com ilhas de es\-ta\-bi\-li\-da\-de, pode se transformar em um mar caótico (compare as figuras \ref{fig:9} e \ref{fig:12}).
Os pontos de equilíbrio instáveis têm essa propriedade devido à existência de separatrizes do movimento (já comentadas anteriormente), que levam ao \emph{caos homoclínico} nessa região, muito mais sensível à perturbação do que o caos gerado por ressonâncias entre as frequências na\-tu\-rais do sistema \cite{deaguiarLivro, tabor1989chaos}. Esse fenômeno também é co\-nhe\-ci\-do como `quebra da separatriz' \cite{lichtenbergLieberman1992}.

\section{Pêndulo simples forçado}
\label{sec:pendulo}

Como um exemplo concreto de aplicação, con\-si\-de\-ra\-mos o pêndulo simples sob a ação da gravidade, formado por uma massa pontual $m$ presa à ponta de uma haste rígida de lado $\ell$ e massa desprezível. A outra ponta da haste está fixa de modo que o pêndulo pode librar (oscilar) ao redor desse ponto de fixação.

A posição do pêndulo pode ser representada pelo ângulo $\theta$ que a haste faz com a vertical, sendo $\theta=0$ a posição de repouso. O movimento pode então ser des\-cri\-to por apenas uma equação para $\theta(t)$, que é obtida aplicando a segunda lei de Newton ao sistema e considerando apenas a aceleração tangencial, na direção do movimento circular. O resultado é  \cite{marion2013classical}
\begin{equation}\label{eq:eqmovPendulo}
\ell\, \ddot{\theta} + g\,\text{sen}\,\theta=0\,,
\end{equation}
onde $g$ é a aceleração da gravidade.
Temos, associada ao sistema pêndulo-Terra, a energia potencial gravitacional (a menos de uma constante arbitrária)
\begin{equation}\label{eq:energpotPendulo}
U(\theta) = - mg\,\ell\cos\theta\,.
\end{equation}

Assim, vemos das equações acima que o movimento tem características de um oscilador na variável $\theta$, ao menos para e\-ner\-gias $E<mg\,\ell$, isto é, tais que a barra não é capaz de atingir  $\theta = \pm\pi$ (que é um ponto de equilíbrio instável). Para pequenos desvios do equilíbrio (estável) $\theta=0$, temos que a aproximação $\text{sen}\,\theta\approx\theta$ na equação (\ref{eq:eqmovPendulo}), ou $\cos\theta\approx 1 - \theta^2/2$ na equação (\ref{eq:energpotPendulo}) é boa e o sistema se comporta como um oscilador harmônico de frequência $\omega_0 = \sqrt{g/\ell\,}$. Conforme consideramos órbitas com maior amplitude angular, começam a ficar importantes também termos de ordem superior nas expansões de $\text{sen}\,\theta$ em série de potências de $\theta$ na equação~(\ref{eq:eqmovPendulo}). Como a série contém apenas termos ímpares, temos que os termos adicionais que aparecem na equação de movimento são potências ímpares de $\theta$, como no caso do oscilador anarmônico discutido acima (o mesmo pode ser visto na expansão de  $\cos\theta$ na energia potencial, que envolve apenas termos pares nas potências de $\theta$).

\begin{figure}
    \centering
    \includegraphics[width = 0.99\columnwidth]{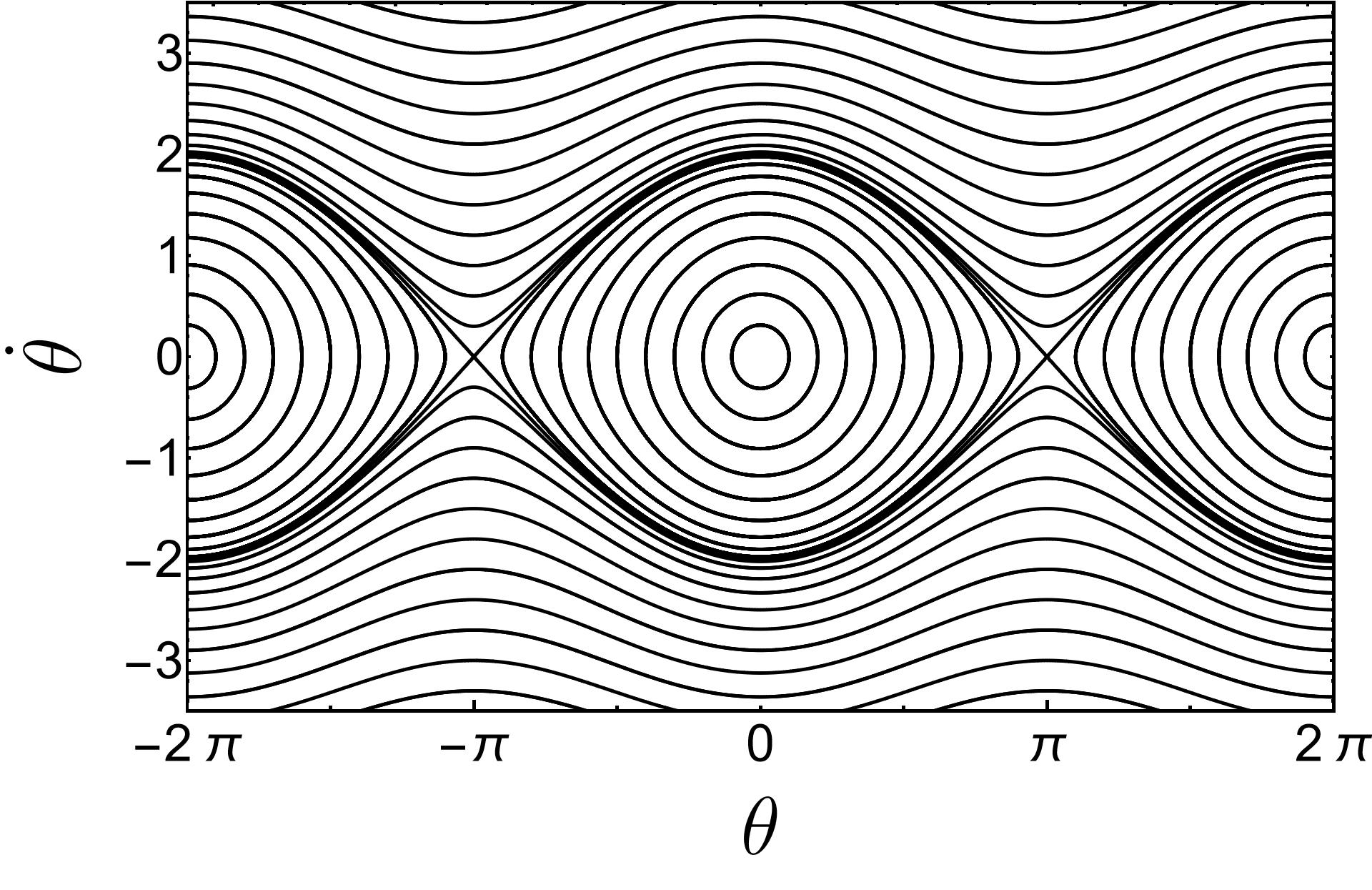}
    \caption{Diagrama de fases para o pêndulo simples, com $g/\ell =1$.}
    \label{fig:13}
\end{figure}

No entanto o pêndulo contém uma característica adicional. Para energias maiores que a do ponto de equilíbrio instável o pêndulo deixa de apenas librar ao redor do ponto de equilíbrio estável e começa a executar rotações completas. O diagrama de fases do pêndulo simples é apresentado na figura \ref{fig:13} (onde fisicamente identificamos $\theta=-\pi$ com $\theta=\pi$ e assim por diante, a cada intervalo de $2\pi$), e as órbitas que dividem essas duas regiões são as separatrizes do sistema (que ``dividem'' os dois tipos de movimento qualitativamente distintos, oscilações ao redor do ponto de equilíbrio estável ou rotações completas da barra).

\begin{figure}
    \centering
    \includegraphics[width = 1\columnwidth]{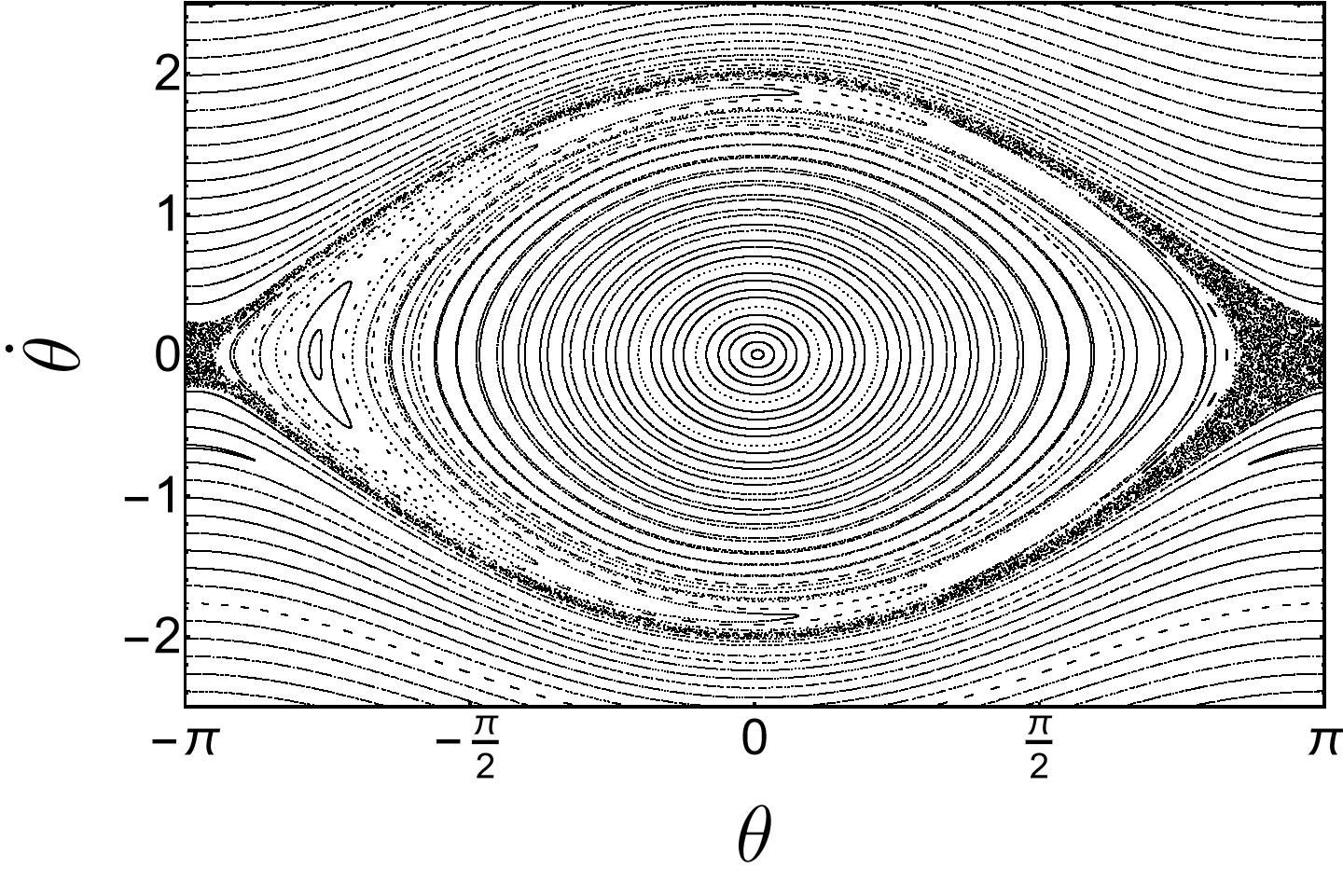}
    \caption{Órbitas do mapa estroboscópico para o pêndulo simples forçado, com $m\,\ell=1$, $g/\ell =1$, $F_0 = 0.01$, $\omega = 2/\pi$. É possível ver a quebra da separatriz em uma região caótica, assim como o surgimento de ilhas de ressonância em forma de banana do lado esquerdo na região de movimento limitado (libração), que para perturbações maiores também se quebrarão em regiões caóticas.}
    \label{fig:14}
\end{figure}

Consideremos agora o caso forçado, com a força externa agindo sempre na direção puramente tangencial ao movimento:
\begin{equation}
 \ddot{\theta} + \frac{g}{\ell}\,\,\text{sen}\,\theta= \frac{F_0}{m\,\ell} \cos(\omega t)\,.
\end{equation}
Vemos da discussão acima e do mapa estroboscópico representado na figura~\ref{fig:14} que o pêndulo engloba dois fenômenos des\-cri\-tos anteriormente: o primeiro é o caos devido a ressonâncias entre as frequências de oscilação do sistema, para variações angulares limitadas próximas do ponto de equilíbrio estável (é possível ver o surgimento de ilhas de ressonância em forma de banana no lado esquerdo da figura \ref{fig:14} que, como vimos, levarão a regiões caóticas para amplitudes maiores da perturbação). Já o segundo é a região caótica que tem as mesmas propriedades qualitativas apresentadas no oscilador de Duffing, devido à quebra da separatriz ao redor do ponto de equilíbrio instável. Uma discussão extensa sobre caos no pêndulo simples forçado (e amortecido) pode ser encontrada na referência \cite{gitterman2010chaotic}.

\section{Caos em Sistemas hamiltonianos}
\label{sec:hamiltonianos}




Na seção anterior vimos que osciladores forçados podem apresentar movimento caótico. 
De fato, quando submetemos o oscilador quártico, o oscilador de Duffing e o pêndulo 
simples a forças externas periódicas, o movimento observado no mapa estroboscópico
passa a apresentar regiões caóticas misturadas a regiões regulares. No entanto,
o oscilador harmônico forçado continua a apresentar um mapa estroboscópico bastante 
re\-gu\-lar. Podemos então nos perguntar sob quais condições a perturbação periódica, ou mesmo
outro tipo de perturbação, vai introduzir caos no sistema.

A resposta a essa pergunta, no contexto de sistemas mecânicos, está ligada ao \emph{teorema de 
integrabilidade de Arnold-Liouville}. 
A demonstração desse teorema pode ser encontrada em \cite{deaguiarLivro, tabor1989chaos, arnoldMmcm} e está
além do escopo deste artigo. No entanto, podemos dar uma ideia de seus elementos principais.  
O primeiro passo para isso é re\-es\-cre\-ver as equações de Newton, que são de segunda ordem no 
tempo, em termos de equações de primeira ordem, conhecidas como equações de Hamilton. 

Vamos considerar uma partícula de massa $m$ se movendo em uma dimensão, sobre a qual age 
uma força conservativa $F$, tal que 
$F(x) = - dU/dx$ onde $U(x)$ é a e\-ner\-gia potencial do sistema. Para o oscilador harmônico, $F=-kx$ e
$U(x) = k x^2/2$. O momento linear da partícula é $p_x = m \dot{x}$ e a segunda lei de Newton pode ser 
escrita como
\begin{equation}
 \dot{p}_x = -dU/dx\, . 
\end{equation}
Juntamente com a relação 
\begin{equation}
    \dot{x} = p_x/m
\end{equation}
temos um sistema de duas variáveis, $x$ e $p_x$, regido por equações diferenciais de primeira ordem no tempo. A energia do sistema
\begin{equation}
 E = m \dot{x}^2/2 + U(x)    
\end{equation}
pode também ser escrita em função de $x$ e $p_x$ como 
\begin{equation}
 H(x,p_x) = p_x^2/2m + U(x)    
\end{equation}
e é chamada de \emph{função hamiltoniana} do sistema. Em termos de $H$, as equações de movimento podem ser escritas como
\begin{eqnarray}
\dot{x} &=& \partial H/ \partial p_x\,, \nonumber \\
\dot{p}_x &=& - \partial H/\partial x\,,    
\label{eqham1}
\end{eqnarray}
que são conhecidas como \emph{equações de Hamilton}. Essas equações são totalmente equivalentes às equações 
de Newton, mas trazem algumas vantagens por serem de primeira ordem.

A teoria por trás das equações de Hamilton é bastante extensa e não entraremos nela com profundidade
neste trabalho. Por enquanto basta dizer que ela se estende a mais dimensões e para mais partículas, assim como as equações de Newton. Em duas dimensões, em particular, as equações de Hamilton são dadas por
\begin{eqnarray}
\dot{x} &=& \partial H/\partial p_x \,, \nonumber \\
\dot{y} &=& \partial H/\partial p_y \,, \nonumber \\
\dot{p}_x &=& - \partial H/\partial x \,, \nonumber \\
\dot{p}_y &=& - \partial H/ \partial y \,,
\label{eqham2}
\end{eqnarray}
onde a função hamiltoniana $H(x,y,p_x,p_y) = p_x^2/2m + p_y^2/2m + U(x,y)$ é a energia do sistema, que permanece constante durante o movimento.
O teorema de integrabilidade de Arnold-Liouville diz que se existir uma outra função $G(x,y,p_x,p_y)$, independente de $H$, que também permanece constante durante o movimento, então a dinâmica não apresenta caos. Para $D$ graus de liberdade ($D=2$ para um oscilador se movendo em duas dimensões, por exemplo) são necessárias $D$ cons\-tan\-tes do movimento independentes, i.e., $D$ funções das coordenadas e dos momentos que permaneçam cons\-tan\-tes durante o movimento, e tais que uma não possa ser escrita em função das outras. O aparecimento de caos está li\-ga\-do à quebra de integrabilidade, isto é,  à falta de cons\-tan\-tes de movimento em número suficiente.

Um exemplo importante é o movimento de uma partícula em um campo central. Nesse caso o vetor momento angular $\vec{L}$ é conservado e o movimento ocorre no plano perpendicular a $\vec{L}$. Como a
energia $H$ e o módulo do momento angular $L$ são constantes, o movimento é regular, sem caos. Por exemplo, para o potencial gra\-vi\-ta\-cio\-nal kepleriano (de uma massa pontual, $- GM/r$) as trajetórias são elipses ou hipérboles. No entanto, se um terceiro corpo for acrescentado ao sistema, ele deixará de ser integrável e irá apresentar regiões de caos. Considere, por exemplo, um asteroide em órbita ao redor do Sol. A presença de Júpiter vai perturbar sua órbita e introduzir caos e instabilidades que podem ser observadas. De fato, o cinturão de asteroides entre Marte e Júpiter apresenta falhas nas regiões caóticas, pois as órbitas desses as\-te\-roi\-des acabaram por se desviar do cinturão. O mesmo acontece em relação às falhas nos anéis de Saturno.

\subsection{Caos em osciladores forçados}

Mas, afinal, o que a dinâmica hamiltoniana e o teorema de integrabilidade têm a ver com nossos osciladores unidimensionais perturbados? Para estabelecer essa 
conexão notamos, em primeiro lugar, que a dinâmica desses osciladores também pode 
ser descrita pela teoria de Hamilton por meio de uma função hamiltoniana dependente do tempo
\begin{equation}
H(x,p_x,t) = p_x^2/2m + U(x) - F_0 x \cos(\omega t)\,.    
\label{hamil1}
\end{equation}
De fato, usando as equações (\ref{eqham1}) vemos que
\begin{eqnarray}
\dot{x} &=& p_x/m\,, \nonumber \\
\dot{p}_x &=& -dU/dx + F_0 cos(\omega t)\,,  \label{eq:oscForxpx}
\end{eqnarray}
ou
\begin{equation}
    \ddot{x} = \frac{\dot{p}_x}{m} =  -\frac{1}{m} \frac{d U}{d x} + \frac{F_0}{m} \cos (\omega t)\,.
    \label{oscfor}
\end{equation}
Veja que $H$ não é constante, pois depende do tempo. 

Vamos agora fazer um truque matemático e transformar essas equações dependentes do tempo para $x$ e $p_x$  em equações para $x$, $p_x$, $y$ e $p_y$ independentes do tempo. Nessa transformação $y$ e $p_y$ são variáveis auxiliares que vamos interpretar a seguir. Definimos  
\begin{equation}
 {\cal H} (x,y,p_x,p_y) = p_x^2/2m + p_y + U(x) - F_0 x \cos(\omega y)\,.
 \label{hamil2}
\end{equation}
Usando as equações (\ref{eqham2}) obtemos
\begin{eqnarray}
\dot{x} &=& p_x/m \label{xx}\,,\\
\dot{y} &=& 1 \label{yy}\,,\\
\dot{p}_x &=& - dU/dx + F_0 \cos (\omega y) \,,\label{pp}\\
\dot{p}_y &=& -F_0 \omega x \sin (\omega y)\,. \label{py}
\end{eqnarray}
A segunda dessas equações mostra que $y=t$. Subs\-ti\-tu\-in\-do esse resultado na terceira equação vemos que (\ref{xx}) e (\ref{pp}) ficam idênticas às equações~(\ref{eq:oscForxpx}), e portanto à equação de movimento original (\ref{oscfor}). Isso mostra que a hamiltoniana do sistema bidimensional (\ref{hamil2}) descreve a mesma dinâmica do sistema (\ref{hamil1}). A equação (\ref{py}) é independente das outras e pode ser resolvida quando $x(t)$ for calculado. Seu valor coincide com $-dH/dt$ (e portanto $p_y = -H$) e reflete o fato de que $H$ não é conservada, embora ${\cal H}$ seja.

Podemos agora aplicar o teorema de Arnold-Liouville para determinar a existência de caos (ou a sua i\-ne\-xis\-tên\-cia). No caso do oscilador harmônico forçado, $U(x) = m \omega_0^2 x^2/2$, de fato existe uma outra constante de movimento ${\cal G}$, diferente de ${\cal H}$, que torna o sistema integrável, sem caos.
Ela é dada por \cite{leach1980complete}
\begin{equation}
    {\cal G} = -[x+g(y)] \sin y - [p_x-h(y)] \cos y
\end{equation}
onde 
\begin{equation}
    g(y) = \frac{\omega_0 F_0}{\omega^2 - \omega_0^2}\, [\cos(\omega y) - \cos (\omega_0 y)]
\end{equation}
e
\begin{equation}
    h(y) = \frac{F_0}{\omega^2 - \omega_0^2}\, [\omega \sin(\omega y) - \omega_0 \sin (\omega_0 y)]\,.
\end{equation}
O leitor pode verificar, usando as equações de movimento e alguma álgebra, que, de fato, $\dot{{\cal G}}=0$.  

Esse exemplo mostra o quão complicado pode ser encontrar as constantes de movimento! Quando mo\-di\-fi\-ca\-mos o potencial harmônico acrescentando termos não li\-ne\-a\-res, a integrabilidade é {\it quebrada}, similarmente ao que acontece quando incluímos Júpiter no sistema asteroide-Sol. A função ${\cal G}$ deixa de ser constante e apenas ${\cal H}$ se mantém fixa durante o movimento. Isso, no entanto, não basta, e trajetórias caóticas aparecem.

\subsection{A dimensão da região disponível para a trajetória no espaço de fases}

Resta então responder a uma pergunta, que unificaria toda a discussão feita até agora. 
Como verificar se uma hamiltoniana conservativa de 2 graus de liberdade, do tipo (\ref{hamil2}), isto é, ${\cal H} (x,y,p_x,p_y)$, admite órbitas caóticas? Vimos acima que órbitas caóticas são aquelas que não admitem duas constantes de movimento independentes. Como a hamiltoniana (energia) é conservada, o problema todo se reduz a verificar se todas as órbitas do sistema admitem uma segunda constante de movimento independente de ${\cal H} (x,y,p_x,p_y)$.

Procurar uma expressão analítica para essa segunda constante de movimento é uma tarefa hercúlea (e até mesmo impossível caso não exista no espaço todo). Assim, a maioria das técnicas para detecção de caos se baseia em métodos numéricos, que permitem enxergar ``assinaturas'' dessa quebra da segunda constante. Voltaremos a isso depois; primeiro, vejamos que tipo de restrição as constantes de movimento impõem às trajetórias. Para isso, supomos que o movimento seja limitado em todas as coordenadas do espaço de fases $(x,y,p_x,p_y)$.

O espaço de fases de um sistema com 2 graus de liberdade tem 4 dimensões. Em princípio, a partícula tem disponível para se movimentar todo esse espaço. No entanto, sabemos que cada constante de movimento pode ser escrita da forma $F(x,y,p_x,p_y) = 0$, de maneira que forma um vínculo entre as coordenadas no espaço de fases. Na prática, isso significa que a partícula não tem mais disponível toda a região, mas sua trajetória precisa obedecer esse vínculo. Sabemos que um vínculo desse tipo corresponde a uma superfície de dimensão 3 em $\mathbb R^4$ \cite{Guidorizziv26e}. Assim, se a partícula possui uma constante de movimento, como a hamiltoniana ${\cal H}$ por exemplo, sua trajetória está restrita a uma superfície de dimensão 3. 

\begin{figure}
    \centering
    \includegraphics[width = 1\columnwidth]{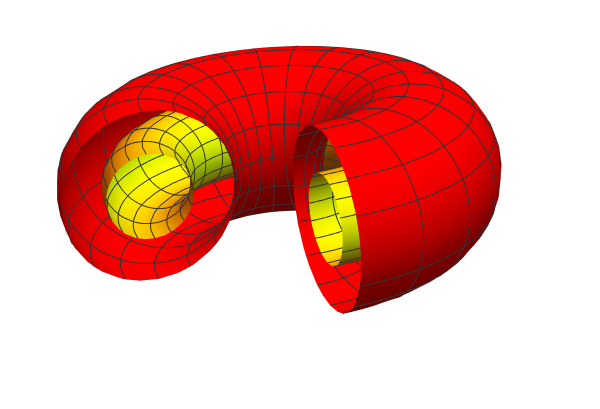}
    \\ \hfill \\
    \includegraphics[width = 1\columnwidth]{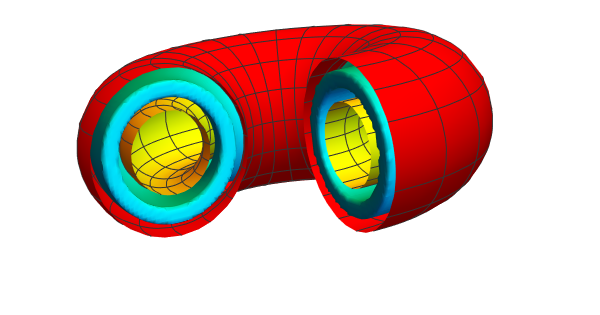}
    \caption{\emph{Acima:} toros (cortados) representando as regiões disponíveis para uma órbita regular percorrer; cada órbita fica restrita a uma superfície toroidal bidimensional (em geral distorcida) no espaço de fases quadridimensional.
    \emph{Abaixo:} ``quebra'' dos toros, correspondendo à falta de uma segunda cons\-tan\-te de movimento; uma órbita caótica percorre toda uma região tridimensional (azul) no espaço de fases quadridimensional.}
    \label{fig:15}
\end{figure}

Temos agora dois cenários possíveis; no primeiro, a partícula está sujeita a uma segunda constante de movimento independente, isto é, a mais um vínculo independente $G(x,y,p_x,p_y) = 0$. Como agora há dois vínculos, o espaço disponível para o movimento da partícula se reduz a uma superfície de dimensão 2 (é possível mostrar que essas superfícies têm o formato de toros  distorcidos \cite{deaguiarLivro, lemos2013mecanica}, ver o painel superior da figura~\ref{fig:15}). Nessa situação dizemos que o movimento é \emph{regular}; de fato, é possível mostrar que se a segunda cons\-tan\-te de movimento é global, é possível resolver o movimento da partícula analiticamente em coordenadas apropriadas, as chamadas \emph{variáveis de ação-ângulo} \cite{deaguiarLivro, lemos2013mecanica, tabor1989chaos}; nessas coordenadas, a trajetória corresponde a um movimento rotacional uniforme ao longo dos dois ângulos de um toro geométrico como o da figura~\ref{fig:15}. 

Já no segundo cenário, dito caótico, a segunda cons\-tan\-te de movimento é ``quebrada'' e a única grandeza conservada é a hamiltoniana ${\cal H} (x,y,p_x,p_y)$; o movimento então ocorre, genericamente, ao longo de uma superfície de dimensão 3 e não pode ser resolvido analiticamente. Entre ou\-tras propriedades, uma órbita caótica genérica preenche densamente uma região tridimensional do espaço de fases (como ilustrado no painel inferior da figura~\ref{fig:15}), o que caracteriza a imprevisibilidade do movimento dada um incerteza nas condições iniciais. Note que, no caso regular, incertezas na condição inicial correspondem a órbitas em toros bidimensionais ligeiramente diferentes, com frequências angulares diferentes mas muito próximas. Agora, uma única órbita preenche toda uma região que seria ``folheada'' por uma família de toros, de modo que a longo prazo duas órbitas com condições iniciais próximas terão percorrido cada uma todo o vo\-lu\-me 3D disponível. Uma dimensão a mais para o movimento implica então em uma imprevisibilidade enorme para tempos longos, se comparada ao caso do movimento restrito a uma superfície bidimensional. É importante notar que, mesmo nesse cenário, algumas trajetórias ainda se comportam como se fossem regulares, percorrendo uma superfície de dimensão dois, como ilustrado nas figuras \ref{fig:12} e \ref{fig:14}. A existência dessas soluções regulares é garantida pelo famoso teorema KAM, de Kolmogorov, Arnold e Moser \cite{arnoldMmcm}. Elas dominam o espaço de fases quando a perturbação é pequena, mas vão sendo substituídas por regiões caóticas conforme a perturbação aumenta.

O mecanismo que leva ao movimento caótico está associado à existência de órbitas periódicas instáveis nessa região 3D e aos chamados {\it emaranhados homoclínicos}. Essa discussão, que está além do escopo deste artigo, pode ser encontrada, por exemplo, nas referências \cite{deaguiar1994RBEF,deaguiarLivro}.

\subsection{Seções de Poincaré}

Existem diversos métodos numéricos que ressaltam ca\-rac\-te\-rís\-ti\-cas diferentes das órbitas e usam essas ca\-rac\-te\-rís\-ti\-cas para classificá-las em regulares ou caóticas, como por exemplo os expoentes de Lyapunov \cite{tabor1989chaos, binneytremaineGD, benettinEtal1980Mecc2} e a análise de frequências da órbita no espaço recíproco \cite{tabor1989chaos, powellPercival1979JPhA, michtchenkoVieiraEtal2018AA}. Queremos, no entanto, apresentar brevemente o método das \emph{seções de Poincaré}, que está intimamente ligado aos mapas estroboscópicos dos osciladores forçados. 

Caso queiramos visualizar a diferença mencionada acima entre o movimento regular restrito a uma superfície de dimensão 2 e o movimento caótico preenchendo um volume de dimensão 3, teríamos que considerar um espaço de fases com 4 dimensões e nele desenhar o vo\-lu\-me 3D correspondendo a um vínculo de energia $\cal H$ cons\-tan\-te. Daí plotaríamos duas órbitas com mesma energia $\cal H$, uma delas preenchendo uma região tridimensional e outra uma região bidimensional. É e\-vi\-den\-te que essa cons\-tru\-ção rapidamente fica impossível de se a\-com\-pa\-nhar na folha de papel ou na tela, se é que é possível começá-la: como desenhar um espaço de fases quadridimensional?

\begin{figure}
    \centering
    \includegraphics[width = 1\columnwidth]{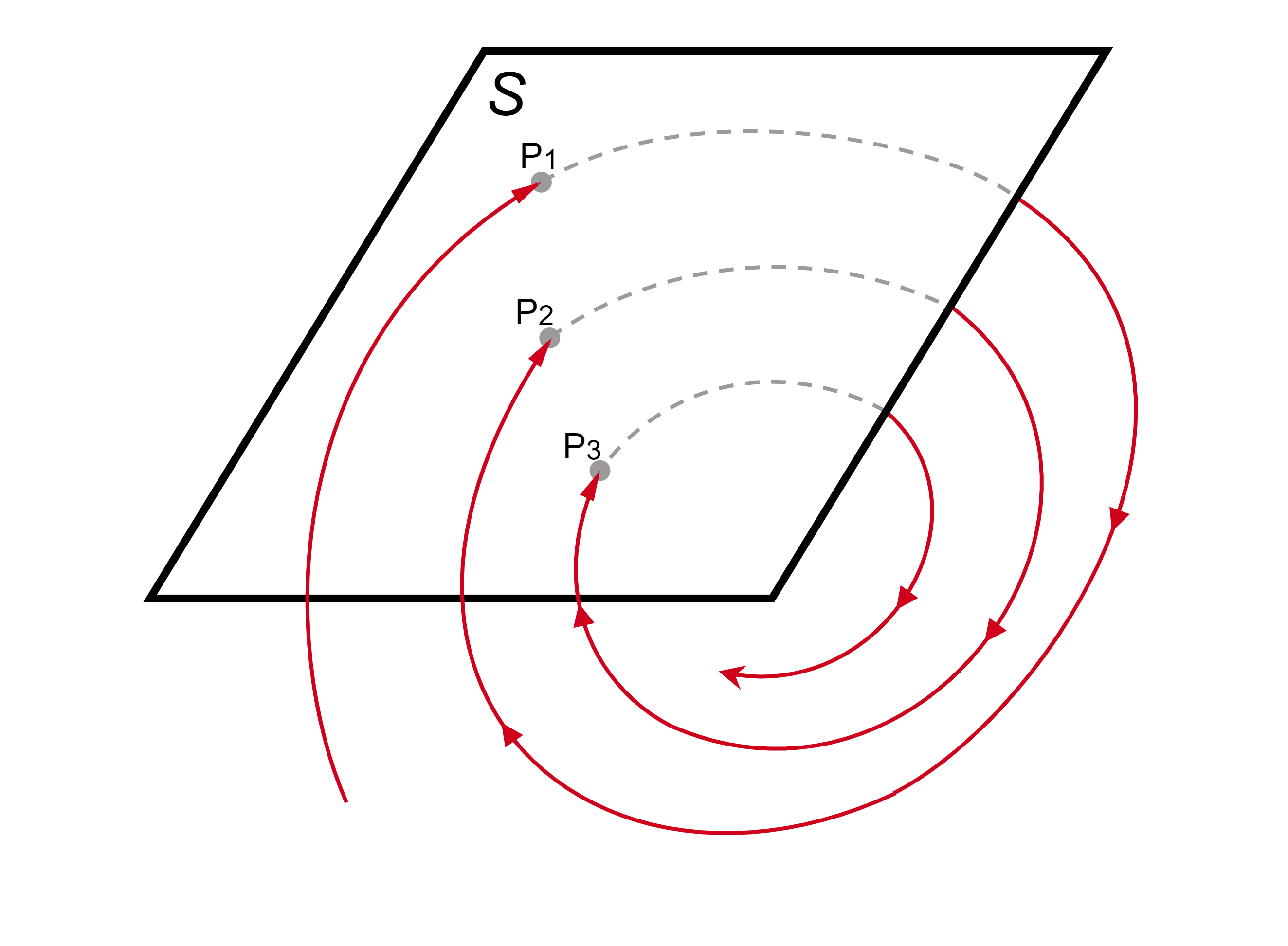}
    \caption{Esquema representando a seção de Poincaré $S$ dentro da superfície 3D de energia constante. Uma única órbita é representada em vermelho, e cada vez que essa órbita cruza a seção (bidimensional) um ponto é plotado na interseção, formando uma sequência $\{ P_1, P_2, P_3, ..., P_n, ...\}$.}
    \label{fig:16}
\end{figure}

A maneira de contornar essa dificuldade vem por meio da técnica das \emph{seções de Poincaré}. Vamos considerar um sistema com 2 graus de liberdade e uma superfície de e\-ner\-gia 3D fixa ${\cal H} = E$. Agora vamos considerar um novo ``vínculo fictício'', não associado a nenhuma constante física e que não influencia no movimento, que terá o papel de permitir enxergarmos uma ``fatia'' dessa superfície de e\-ner\-gia 3D no papel. Esse novo vínculo fictício pode ser da forma $y=0$, por exemplo. Assim, fixamos uma superfície fictícia bidimensional no espaço de fases dada pelas condições ${\cal H} (x,y,p_x,p_y) = E$ e $y=0$. Ao longo dessa superfície fictícia, a \emph{seção de Poincaré}, três coordenadas do espaço de fases variam (mas não de maneira independente): $x$, $p_x$ e $p_y$. Além disso, em geral, considerando apenas as órbitas com ${\cal H} = E$, toda a vez que uma órbita atingir $y=0$ ela marcará um ponto na seção. Sabemos que a partícula vai e volta na coordenada $y$, marcando um ponto cada vez; assim, para que possamos enxergar a órbita como se ``costurasse'' a seção de Poincaré, pegamos o cruzamento em apenas um sentido, o que pode ser feito escolhendo sempre $p_y>0$ para marcar os pontos (ver figura~\ref{fig:16} para uma ilustração). Podemos então representar essa superfície no plano $x$--$p_x$, uma vez que dado um ponto nesse plano fica subentendido que o ponto na verdade pertence à superfície curva no espaço de fases determinada por $y=0$ e por ${\cal H} = E$, e que $p_y$ é obtido por meio dessas duas condições.

Chegamos então à seguinte situação: temos uma superfície bidimensional no espaço de fases quadridimensional, a seção de Poincaré. Essa superfície pode ser projetada em um plano dado por duas coordenadas, neste caso $x$ e $p_x$. Cada órbita no espaço de fases será então representada por uma sequência de pontos nesse plano (que representam as interseções da órbita com a seção). A pergunta que fica é: é possível relacionar as propriedades acima que determinam as dimensões disponíveis para o movimento com propriedades dessas sequências de pontos ao aparecerem no plano $x$--$p_x$? A resposta, como esperado, é sim. Tomemos uma órbita regular, aquela cuja partícula se movimenta sobre a superfície fechada (toroidal) de dimensão 2. A interseção dessa superfície com a seção de Poincaré nos dá então uma seção (uma ``fatia'') desse toro distorcido, que então é vista como uma curva fechada no plano $x$--$p_x$. Se a órbita preenche densamente o toro distorcido para tempos muito longos, os cruzamentos com a seção de Poincaré preenchem toda a curva fechada; se a quantidade de pontos plotada é grande o suficiente, o que vemos é que órbitas regulares são representadas por curvas fechadas na seção de Poincaré (veja a órbita azul da figura~\ref{fig:10}). 

Agora, para uma órbita caótica, ela preenche densamente uma região tridimensional do espaço de fases. A interseção dessa região tridimensional com a seção de Poincaré é uma região bidimensional, ou seja, uma área no plano $x$--$p_x$. Em outras palavras, a irregularidade da órbita no espaço de fases se traduz no preenchimento denso de uma área na seção de Poincaré (órbitas vermelha e verde da figura~\ref{fig:10}). Em geral, fixado o vínculo ${\cal H} = E$, existirão tanto órbitas regulares quanto órbitas caóticas representadas nas seções de Poincaré, como comentamos anteriormente.

E qual a relação disso com os osciladores forçados? Como já vimos, definindo $y$ e $p_y$ da forma~(\ref{hamil2}) e comparando com o sistema físico (\ref{hamil1}), vemos que $y$ (ou $t$) entra nas equações de movimento apenas na forma de um ângulo; isto é, matematicamente o ponto é o mesmo caso façamos $\omega y' = \omega y + 2\pi$. Assim, é como se $y$ morasse sobre o círculo, sendo uma coordenada angular. A condição $y=0$ da seção de Poincaré então significa na verdade que $\omega y=0$ no círculo, isto é, a menos de um múltiplo de $2\pi$. Em outras palavras, os pontos na seção de Poincaré serão dados por $y_n = 2 n\pi/\omega$ ou, em termos do tempo físico $t$, $t_n = 2 n\pi/\omega$. E isso é justamente a condição que utilizamos para construir os mapas estroboscópicos: as seções estroboscópicas no plano $x-\dot{x}$ são então justamente seções de Poincaré do sistema hamiltoniano (\ref{hamil2}) quando fazemos a identificação das coordenadas do espaço de fases com as do sistema físico (\ref{hamil1}). Consequentemente, a ideia intuitiva de caos apresentada para os osciladores forçados pode ser vista como um exemplo da formulação mais rigorosa apresentada nesta seção.

\section{Conclusões}
\label{sec:Conclusoes}

Embora a teoria da dinâmica não-linear e sua relação com caos e integrabilidade em sistemas hamiltonianos conservativos seja um assunto extenso, tão extenso quanto necessário for se aprofundar, o conhecimento de mecânica clássica básica se mostra suficiente para um primeiro contato com algumas características que não aparecem em sistemas lineares.
Vimos que é possível, por meio das técnicas adequadas, partir gradualmente do caso bem conhecido do oscilador harmônico forçado para a análise do movimento de osciladores não-lineares na presença de forças externas periódicas no tempo, algo bastante mais delicado e que foge do ``kit de ferramentas'' usual aprendido nos cursos de graduação em física. Os mapas estroboscópicos permitem fazer uma distinção visual, qualitativa, dos tipos de movimento (regular e caótico) de maneira bastante intuitiva, tendo como base sua aplicação ao oscilador harmônico forçado.

Como mencionado ao longo do texto, um `caminho' para se aprofundar em sistemas hamiltonianos conservativos é o estudo de outros artigos da Revista Brasileira de Ensino de Física, como por exemplo \cite{deaguiar1994RBEF}, seguindo por estudos sistemáticos por meio de livros sobre o assunto como \cite{lemos2013mecanica, deaguiarLivro, goldstein2002classical} e depois por livros especializados da área \cite{tabor1989chaos, lichtenbergLieberman1992}, assim como por material aplicado a di\-fe\-ren\-tes áreas da física e astrofísica, como por exemplo \cite{binneytremaineGD, contopoulosOCDA2002, wimberger2014nonlinear, huang2018relativistic, theoryoforbits2, arnoldMmcm}.

Não comentamos aqui sobre o caso em que há dissipação de energia, isto é, o caso em que a equação de movimento do oscilador tem um termo de ``arraste'' proporcional a $\dot x$, por exemplo. Nesse caso, embora a técnica dos mapas estroboscópicos ainda possa ser utilizada, a teoria por trás do aparecimento de caos e a interpretação das figuras são diferentes \cite{alligood2000bookchaos, strogatz2018nonlinear, ott2002chaos, lichtenbergLieberman1992, gitterman2010chaotic}. Em particular, não há ``conservação de área'' nas seções de Poincaré, devido à dissipação de energia; isso dá origem por exemplo a estruturas fractais no espaço de fases que descrevem o estado do movimento para tempos muito longos, os \emph{atratores estranhos}, levando a toda uma gama de fenômenos inexistentes em sistemas conservativos \cite{alligood2000bookchaos, ott2002chaos, lichtenbergLieberman1992}. Convidamos o leitor interessado a consultar livros es\-pe\-cia\-li\-za\-dos, em parte citados aqui.

\subsection*{Agradecimentos}

Este trabalho recebeu apoio financeiro da Fundação de Amparo à Pesquisa do Estado de São Paulo (FAPESP) sob os processos n$^o$ 2023/09170-8 (L.H.R.D) e n$^o$ 2021/14335-0 (M.A.M.A.). Também foi parcialmente financiado pelo Conselho Nacional de Desenvolvimento Científico e Tecnológico, CNPq, processos 800566/2022-0 (L.H.R.D) e 301082/2019-7 (M.A.M.A.).

%

%


%





\end{document}